\documentclass[12pt,letterpaper]{article}

\usepackage{jheppub}
\usepackage{color}
\usepackage[dvipsnames]{xcolor}
\usepackage{graphicx}
\usepackage{wrapfig, enumerate, slashed}

\usepackage[noabbrev]{cleveref}

\usepackage{footmisc}
\usepackage{amsmath}
\usepackage{wasysym} 
\usepackage{graphicx}
\usepackage{color}
\usepackage{comment}
\usepackage{hyperref}

\newcommand{\bee}{\begin{eqnarray}}
\newcommand{\eee}{\end{eqnarray}}
\newcommand{\beeq}{\begin{equation}}
\newcommand{\eeeq}{\end{equation}}

\renewcommand{\vec}{\bf}

\newcommand{\iu}{{i\mkern1mu}}

\newcommand{\beqa}{\begin{eqnarray}}
\newcommand{\eeqa}{\end{eqnarray}}
\newcommand{\be}{\begin{equation}}
\newcommand{\ee}{\end{equation}}
\newcommand{\ba}{\begin{array}} 
\newcommand{\ea}{\end{array}}
\def\bea{\begin{eqnarray}}
\def\eea{\end{eqnarray}}
\newcommand{\nn}{\nonumber}

\newcommand{\Dfb}{\mathord{\buildrel{\lower3pt\hbox{$\scriptscriptstyle{\leftrightarrow \tiny{ \ \ \ } }$}}\over {D^{\mu}}}}

\newcommand{\Dfbd}{\mathord{\buildrel{\lower3pt\hbox{$\scriptscriptstyle\leftrightarrow$}}\over {D}_{\mu}}}

\usepackage{bbold}
\usepackage{comment}

\begin{document}

\title{One Loop Thermal Effective Action}

\author{Joydeep Chakrabortty,}
\author{Subhendra Mohanty}

\affiliation{Indian Institute of Technology Kanpur, Kalyanpur, Kanpur 208016, Uttar Pradesh, India}

\emailAdd{joydeep@iitk.ac.in}
\emailAdd{mohantys@iitk.ac.in}

\abstract{ We compute the one loop effective action for a Quantum Field Theory at finite temperature, in the presence of background gauge fields,  employing the Heat-Kernel method. This method enables us to compute the thermal corrections to the Wilson coefficients associated with effective operators up to arbitrary mass dimension, which emerge after integrating out heavy scalars and fermions from a generic UV theory. The Heat-Kernel coefficients are functions of non-zero background  `electric', `magnetic' fields, and Polyakov loops. A major application of our formalism is the calculation of the finite temperature Coleman-Weinberg potential  in effective theories, necessary for the study of phase transitions.  A novel feature of this work is the systematic calculation of the dependence of Polyakov loops on the thermal factors of Heat-Kernel coefficients and the Coleman-Weinberg potential.  We study the effect of Polyakov loop factors on phase transitions and comment on future directions in applications of the results derived in this work. 
}

\maketitle

\section{Introduction}

The determination of the nature of the  electroweak phase transition is of great interest as a first order phase transition (FOPT)  would be useful for explaining electroweak baryogenesis \cite{Anderson:1991zb,Trodden:1998ym,Cline:2006ts}. A FOPT in the early universe will also produce stochastic gravitational waves, which may be observable in Earth or Space-based gravitational wave detectors or Pulsar-Timing Arrays. The frequency spectrum of the GW signals is related to the temperature of the FOPT; therefore, particle physics models' phase transitions are directly testable in GW observations \cite{Caprini:2009yp}.

In the Standard Model lattice calculations have shown that the electroweak phase transition is of first order for Higgs mass $m_H \lesssim 68 \,{\rm GeV}$ \cite{Kajantie:1996mn, Csikor:1998eu} which implies that in the SM with Higgs mass $m_H \simeq 125 {\rm GeV}$, the EW phase transition is a smooth crossover. 
The addition of dimension-6 operators in a SMEFT gives FOPT for a  Higgs mass as high as ($m_H <170 ~{\rm GeV}$) for even a low cut-off scale  $\Lambda < 800~ {\rm GeV}$ \cite{Grojean:2004xa, Bodeker:2004ws, Delaunay:2007wb,Banerjee:2024qiu} and with the possibility of observations in gravitational wave detectors \cite{Bodeker:2004ws}.  UV extensions of the Standard Model have been constructed, and collider bounds dark matter direct detection bounds on the parameters show the existence of a viable parameter space in the BSM models where a FOPT is achieved by the $125 ~{\rm GeV}$ SM Higgs   \cite{Chung:2012vg, Blinov:2015vma, Cline:2021iff}.

Standard Model Effective Field Theories  (SMEFT) can be used as efficient ways of studying many classes of UV extensions of SM by the addition of higher dimensions operators suppressed by powers of the cut-off scale $\Lambda $, which is of the order of the masses of the heavy fields of the UV theory integrated out. 
While it would be more efficient to study the problem of EW-phase transitions in the SMEFT framework, a comparison of predictions of  UV complete models with those of effective theories shows a mismatch of parameters space for FOPT between the two \cite{Damgaard:2015con,deVries:2017ncy}. It is noted that in the parameter space for achieving FOPT, dimension eight operators become of the same magnitude as dimension six operators \cite{deVries:2017ncy}, which raises the question of the efficacy of using SMEFT studying phase transitions \cite{Postma:2020toi}. There is also an issue of non-decoupling of the heavy field mass parameter in the light field effective potential in cases of heavy-light mixing in the UV theory \cite{Kanemura:2022txx, Hashino:2022ghd, Florentino:2024kkf, Athron:2023xlk}.

Heat-Kernel is one of the finest among the existing methods to compute  one loop effective action \cite{Belkov:1995gjw, Vassilevich:2003xt, Avramidi:2001ns, Banerjee:2023iiv, Chakrabortty:2023yke, Banerjee:2023xak}. This method can be easily extended to accommodate the scenarios where the heavy fields are integrated out, leading to effective action at one loop or beyond. Recently, in a series of papers, the one-loop effective action has been computed up to dimension eight for scalars and fermions
  \cite{Banerjee:2023iiv, Chakrabortty:2023yke, Banerjee:2023xak}. Similar results have been extended beyond one loop in \cite{Jack:1982hf, Banerjee:2024rbc}. All these constructions are done for zero temperature field theory. There are attempts to extend the very similar idea for finite temperature for some specific low energy QCD scenarios \cite{Megias:2003ui}. We highlight in this work that it is possible to establish relations among the zero-temperature Heat-Kernel coefficients (HKCs) with the ones in the presence of finite temperature $(T)$. In the case of $T\neq 0$, the truncation of the asymptotic series in Schwinger parameter $(t)$ is not straightforward as different powers of the parameters mentioned above contribute to the same HKC. We discuss the method to compute $T\neq 0 $ HKCs in great detail up to dimension-6 effective operators, and that can be generalised further if required.  

In order to study phase transitions, one has to calculate the Coleman-Weinberg (CW) effective potential \cite{Coleman:1973jx, Weinberg:1973am}, which is a one-loop effective theory obtained by expanding a scalar field around a fixed background and integrating out the short distance quantum fluctuations. Dolan and Jackiw first worked out the finite temperature corrections to the CW potential \cite{Dolan:1973qd} using the real-time diagrammatic method and the imaginary time path integral formalism. This procedure needs to be augmented by replacing the light field masses with their hard-loop thermal corrections in a procedure known as Daisy resummation, first advocated by Parwani \cite{Parwani:1991gq} and Arnold-Espinosa \cite{Arnold:1992rz}. The thermal resummation for multi-field potentials is carried out in \cite{Bahl:2024ykv}. Finally, the finite-temperature effective theory of the SM is presented as effective theory 3-dimensional theory in \cite{Farakos:1994xh, Kajantie:1995dw, Braaten:1995cm, Brauner:2016fla} using diagrammatic techniques. All these different approaches to the finite temperature effective potential yield quantitatively different results as was surveyed in \cite{Croon:2020cgk, Ekstedt:2024etx}. In this paper, we compute the effective action at finite temperatures by starting from a UV  theory and integrating out the heavy scalar bosons and fermions. The order $m^4$ corrections give the free energy, and this is used for computation of the thermal Coleman-Weinberg effective potential for light scalars -like the standard model Higgs. The next order, $m^2$ correction, gives the thermal mass. We compute the thermal corrections to the effective action at each order of expansion in the heavy mass scale $1/m$ to consistently match the order of the effective action term. We illustrate our results by computing the thermal Coleman-Weinberg potential of the Higgs arising from an integrated out heavy scalar and discuss its implication in phase transitions.
We also give the derivation of the finite temperature non-Abelian generalisation of the Euler-Heisenberg effective action of QED \cite{Dunne:2004nc} by integrating out heavy fermions. This effective action has applications in quark-gluon plasma \cite{Megias:2003ui}, in the calculation of the Schwinger mechanism \cite{Sachs:1991en} and the Casimir effect \cite{Hawking:1976ja, Blau:1988kv}.

A novel feature of the effective potentials obtained from integrating out heavy gauged fields is the emergence of the Polyakov loop \cite{Polyakov} contribution to the effective potential. The Polyakov loop (PL), defined as $\Omega=\mathbb{T}\,exp\{- \int_0^\beta A_0 (x_0,x_i) dx_0\}$ was first introduced in the context of QCD as a useful order parameter that distinguishes the confined phase from the deconfined phase of quarks at finite temperature \cite{Polyakov, Svetitsky}. Physically, it represents the free energy (or equivalently, pressure) associated with the unconfined quarks in a thermal bath of gluons. In a general SU(N) gauge theory, the contribution Polyakov loop in the effective action up to quartic order in $\Omega$ was computed by Weiss \cite{Weiss:1980rj}. In the context of QCD, the PL contribution to the quark-gluon pressure was computed by  \cite{Pisarski:2000eq, Pisarski:2002ji}, and a calculation of a background Polyakov loop potential using functional integrals was carried out in \cite{Haas:2013qwp, Pasechnik:2023hwv}. The contribution to phase transitions from Polyakov loops has been studied in the context of composite Higgs models \cite{Fujikura:2023fbi}, dark sector non-Abelian gauge theories \cite{Pasechnik:2023hwv} and in the QCD phase transition \cite{Shao:2024dxt}. In this paper, using the Heat-Kernel method of obtaining the effective action at finite temperature by integrating out heavy bosons and fermions gives us a systematic procedure for including the contribution of Polyakov loop terms to all orders. We then study the implications of Polyakov loops for phase transitions.

This paper is organised as follows. In Section~\ref{EFT}, we introduce the Heat-Kernel method and use it for calculating the effective action at finite temperatures by integrating out heavy scalars in the UV theory. In Section~\ref{sec:fermion}, we derive the expressions for the finite temperature effective action resulting from integrating out fermions. In Section~\ref{CW-T2}, we give the general procedure for the consistent calculations of the Coleman-Weinberg effective potential in effective theories, and we illustrate this with a simple example of calculation of CW potential on integrating out a heavy scalar. In Section~\ref{PhaseT}, we discuss the implications of our results in phase transitions. In Section~\ref{PL}, we discuss the contributions of Polyakov loops to the effective potential and study their implication for phase transitions. In Section~\ref{Conclusions}, we summarise our results and point to future applications. In Appendix~\ref{Appendix}, we give the detailed steps of derivations and list all the Heat-Kernel with their associated thermal factors used in the paper. 


\section{Finite temperature Effective Action}\label{EFT}
At zero temperature, we define the quantum field theory in $d+1=4$ dimensional space-time. 
The effective action for a theory having a strong elliptic operator of the form $\Delta = D^2+m^2+U$, where $m$ is the mass term and $U=\frac{\delta^2 \mathcal{L}}{\delta \phi^{\dagger} \phi}$ captures the interaction of the theory, is given in terms of Heat-Kernel coefficients: $b_i$  \cite{Belkov:1995gjw, Vassilevich:2003xt, Avramidi:2001ns, Banerjee:2023iiv, Chakrabortty:2023yke, Banerjee:2023xak}
\begin{eqnarray}\label{eq:EFT-b}
     b_0&=& 1, b_1= U, b_2 = U^2 + \frac{1}{6} G_{\mu \nu}^2 + \frac{1}{3} U_{\mu\mu}, \nonumber \\
     b_3&=& U^3-\frac{1}{2} (U_{\mu})^2+\frac{1}{2} U G_{\mu \nu}^2 -\frac{1}{10} (J_\nu)^2 +\frac{1}{15}G_{\mu \nu} G_{\nu \rho} G_{\rho \mu} \nonumber \\
     & & + \frac{1}{10} U_{\nu \nu \mu \mu} -\frac{1}{30} [D_\mu, [D_\nu,[D_\nu,J_\mu]]],
\end{eqnarray}
where $ U_{\mu\mu}= [D_\mu, [D_\mu, U]]$, current $J_\nu = [D_\mu, G_{\mu \nu}]$, and $G_{\mu \nu }=[D_\mu, D_\nu]$ is the field tensor with $D_\mu = I D_\mu^{E}$ . \\
To compute the finite temperature effective action, we will adopt a very similar trajectory as discussed in Ref.~\cite{Megias:2003ui, Moral-Gamez:2011wcb}.
To generalize the effective action computation at finite temperature, i.e., $T\neq 0$, we will start with a $(d+1)$ dimensional Euclidean manifold as $\mathcal{M}_d \times R^1$. Here, $\mathcal{M}_d$ is the $d$-dimensional spatial manifold, and $R^1$ is the additional non-compact time direction. We will compactify this direction using periodic boundary conditions
\be
\Phi(t+\beta, {\vec x}) =\pm\, \Phi(t , {\vec x}),
\ee
where the $+$ sign is for bosons and $-$ sign for fermions and $\beta=1/T$, where $T$ is the temperature. Note that the periodic boundary condition is employed only for the wave functions along the $R^1$ component. Thus, the fields in the $d$-dimensional manifold have no boundary, and thus, no boundary condition is imposed along these directions.
Here, we have $(d+1)$ dimensional Euclidean coordinates: $\mu \in \{i, 0\}$ with $i=1,.., d$ is the space-time manifold coordinates. Similarly, $D_\mu^E = \partial_\mu +A_\mu$ is realised with $\partial_\mu = \{\partial_i , \partial_0\}$, and $A_\mu = \{A_i , A_0\}$. Now, when we consider a background $SU(N)$ gauge theory, $A_\mu = A_\mu^a T^a$ where $a=1, .., (N^2-1)$, and $T^a$ are $SU(N)$ generators in fundamental representation. It is worth mentioning that both $A_i$ and $A_0$ will carry gauge charges, depending on the gauge quantum number of the field being integrated out. 

The working Lagrangian of the real scalar field in $(d+1)$ dimensional Euclidean\footnote{Now onward, we will use all the fields and in Euclidean space and $E$-index will be suppressed.} manifold  is given as 
\begin{equation}
 \mathcal{L}= \phi^{\dagger}(-D^2+m^2+U) \phi,
\end{equation}
where $D^2=D_0^2+D_i^2$ is the covariant derivative in Euclidean space $\equiv (D_\mu)^E (D^{\mu})^E$. 
The effective Lagrangian is given as \cite{Avramidi:2001ns, Megias:2003ui, Banerjee:2023iiv, Chakrabortty:2023yke, Banerjee:2023xak}
\begin{equation}
    \mathcal{L}_{eff}= c_s\;\; \text{tr}\; \int_0^{\infty} \frac{dt}{t} K,
\end{equation}
where $c_s=1/2(1)$, and $ -1/2$ for real(complex) bosons and fermions, respectively. The interaction Lagrangian is related to the other thermodynamic quantities as $ \mathcal{L}_{eff} = -V(\text{potential})=-\mathcal{F} (\text{free-energy})=p (\text{pressure})$. Here, the trace of the HK $K(t;x,x;\Delta)$ is given as \footnote{We are using all negative Euclidean metric.} 
\begin{eqnarray}
    \text{tr} K (t;x,x;\Delta) &\equiv&    \left\langle   x| e^{-t\Delta}|x  \right\rangle 
    = \int \frac{d^{d+1}p}{(2\pi)^{d+1}} e^{-m^2t}  \left\langle x| e^{-(-D^2+U)t}|p \right\rangle    \left\langle p|x \right\rangle\\
    &=& \frac{1}{\beta} \sum_{p_0} \int \frac{d^{d}p}{(2\pi)^d} e^{-m^2t} e^{-[-(D_i+ \iu p_i)^2-Q^2+U)]t}.
\end{eqnarray}
Here, we perform the integration over the additional direction (denoted as $0$) using periodic boundary condition, and that leads to the sum over $p_0$, which is the so-called Matsubara frequency
$p_0=\frac{2\pi n}{\beta}$, with $n \in \mathbb{Z}$. Here, we normalize the zero-momentum state as $\left\langle x|0\right\rangle=1$. We define $Q \equiv D_0 +  \iu  p_0= \partial_0 + A_0 +  \iu p_0$, and the thermal Wilson loop, known as the Polyakov loop, as
\be
 \Omega (x_i)= \mathbb{T} \;exp\{ - \int_{x_0}^{x_0+\beta} A_0 (x_0^{'}, x_i) dx_0^{'}\} ,
\ee
 where $\mathbb{T}$ is the path-ordering.

Here, we work with a modified temporal gauge $D_0 A_0 =0 $, which is a residual one of gauge choice $A_0=0$. For $A_0$ not having explicit temperature, i.e., $x_0$ dependence, Polyakov loop reduces to 
$\Omega (x_i)= \;e^{ - \beta A_0(x_i) } = \;e^{ - \beta D_0 }$, as due to periodicity $\;e^{ - \beta \partial_0 }=1$  \cite{Megias:2003ui}.

Then, 
\begin{eqnarray}
  \text{tr} K (t;x,x;\Delta) &\equiv& \frac{1}{\beta} \sum_{p_0} \int \frac{d^{d}p}{(2\pi)^d} e^{-m^2t} e^{-[-(D_i+ \iu p_i)^2 - Q^2+U)]t}\\
  &=& \frac{1}{\beta} \sum_{p_0} \sum_{k} \frac{1}{(4\pi t)^{\frac{d}{2}}} e^{-m^2t} b_k(U-Q^2, D_i^2) \frac{(-t)^k}{k!}\\
    &=& \frac{1}{\beta} \sum_{p_0} \sum_{k} \frac{1}{(4\pi t)^{\frac{d}{2}}} e^{-m^2t} e^{Q^2t} \tilde{b}_k(U-Q^2, D_i^2) \frac{(-t)^k}{k!}.
\end{eqnarray}
Recall that the zero temperature HKCs, for $d$-dimensional Euclidean manifold, are given as 
\begin{eqnarray}
&& b_0=1, b_1= U, b_2 = U^2 +\frac{1}{6} G_{ij}^2 -\frac{1}{3} U_{ii}, \\
      b_3&=& U^3+\frac{1}{2} (U_{i})^2+\frac{1}{2} U G_{ij}^2 +\frac{1}{10} (J_i)^2 - \frac{1}{15}G_{ij} G_{jk} G_{ki} \nonumber \\
       & & + \frac{1}{10} (U)_{iijj} -\frac{1}{30} [D_i, [D_j,[D_j,J_i]]].
\end{eqnarray}
   
Let us define the thermal wave function as \cite{Megias:2003ui}
\begin{equation}
    \varphi_k(\Omega; t/\beta^2) = (4\pi t)^{1/2} \frac{1}{\beta}\sum_{p_0=\frac{2\pi n }{\beta}} t^{k/2} Q^k e^{Q^2 t}.
\end{equation}
Thus, in the presence of temperature, i.e., $T\neq 0$, $b_0$ can be written in terms of the thermal wave function, for bosonic case,\footnote{In case of fermion this modifies as $\Omega \to -\Omega$. In the next part, we will use it to compute the effective action for the fermions.} as \cite{Megias:2003ui}
\begin{eqnarray}
      \varphi_0(\Omega; t/\beta^2) &=& (4\pi t)^{1/2} \frac{1}{\beta}  e^{Q^2 t} =  \sum_n (\pm \Omega)^n  e^{-n^2 \beta^2/4 t},
\end{eqnarray}
where the $+$ and $-$ signs are for bosonic and fermionic fields, respectively, taking care of periodic and anti-periodic boundary conditions, respectively. We will formulate the prescription for Scalar Quantum Field Theory (SQFT) and then discuss how this can be generalized for fermion fields.\footnote{Here, we define $gA_0\equiv A_0$ and later to revive the effect of $g$ we can replace $Log(\Omega)$ by $\frac{Log(\Omega)}{g}$.}  Here, we have used the following property  \cite{Megias:2003ui}
    \begin{equation}
        \sum_n \mathcal{F} ( \iu p_0+D_0) = \sum_n \mathcal{F} ( \iu p_0 - \frac{1}{\beta} ln\; \Omega).
    \end{equation}
Note that after incorporating the thermal effects, the interaction term $U$ is modified to $(U-Q^2)$, and that is why the $b_k$'s are accordingly modified as follows
\begin{eqnarray}
    b_0 & = & 1, b_1= U-Q^2, b_2 = (U-Q^2)^2 +\frac{1}{6} G_{ij}^2 -\frac{1}{3} (U-Q^2)_{ii},  \\
     b_3&=& (U-Q^2)^3 + \frac{1}{2} [(U-Q^2)_{i}]^2+\frac{1}{2} (U-Q^2) G_{i j}^2 +\frac{1}{10} (J_i)^2 - \frac{1}{15}G_{i j} G_{j k} G_{k i} \nonumber\\
     & &  + \frac{1}{10} (U-Q^2)_{iijj} -\frac{1}{30} [D_i, [D_j,[D_j,J_i]]],
\end{eqnarray}
where, $U_{ij} \equiv [D_i, [D_j, U]]$. Comparing above two equations between $b_k$ and $\tilde{b}_k$, we can express $\tilde{b}_k$ in terms of $b_k$
\begin{eqnarray}
    \sum_{k_1} b_{k_1} \frac{(-t)^{k_1}}{{k_1}!} & = &  e^{Q^2 t} \sum_{k_2} \tilde{b}_{k_2} \frac{(-t)^{k_2}}{{k_2}!}
    = \sum_{k_2,k_3}  \frac{(Q^2 t)^{k_3}}{{k_3}!} \tilde{b}_{k_2} \frac{(-t)^{k_2}}{{k_2}!} \nonumber \\
    &=& \sum_{{k_2,k_3}} (-1)^{k_2} \tilde{b}_{k_2}  \frac{(Q^2)^{k_3}}{{k_3}! {k_2}!} (t)^{k_2+k_3}.
\end{eqnarray}
Now, by matching the coefficients of $t^m$ from both sides, we find $m=k_1 = k_2+k_3$.
We give the finite temperature  HKCs in Appendix~\ref{HKC} explicitly, and we agree with results in Ref.~\cite{Megias:2003ui} depicted in different forms. 

The trace of the Heat-Kernel is computed as  \cite{Megias:2003ui}
\begin{equation}
	\text{tr} K (t;x,x;\Delta) = \frac{1}{\beta} \sum_{p_0} \sum_{k} \frac{1}{(4\pi t)^{\frac{d}{2}}} e^{-m^2t} e^{Q^2t} \tilde{b}_k(U-Q^2, D_i^2) \frac{(-t)^k}{k!}.
\end{equation}
Our primary focus is to compute the effective action up to dimension six terms, i.e., up to $k=3$.
Thus, 
\begin{eqnarray}
	\text{tr} K (t;x,x;\Delta) &\equiv &    \text{tr} K (t;x,x;\Delta) |_{0,1,2}   = \frac{1}{\beta} \sum_{p_0} \frac{1}{(4\pi t)^{\frac{d}{2}}} e^{-m^2t} e^{Q^2t} 
	[ \tilde{b}_0 - \tilde{b}_1 t +  \tilde{b}_2 t^2/2! -  \tilde{b}_3 t^3/3!] \nonumber \\ 
	&= &    \frac{1}{\beta} \sum_{p_0} \frac{1}{(4\pi t)^{\frac{d}{2}}} e^{-m^2t} e^{Q^2t} 
	\Bigg[ \tilde{b}_0 - \tilde{b}_1 t +  \tilde{B}_1 t^2 + Q \tilde{B}_2 t^2 \nn \\ 
	& & - \Big[ \tilde{b}_3^{0} + \tilde{b}_{31}^{T} + Q\;\tilde{b}_{32}^{T}  + Q^2\;\tilde{b}_{33}^{T}
    + Q^2\;\tilde{b}_{34}^{T}\Big] t^3 \Bigg] \nonumber \\
	&= &    \frac{1}{\beta} \frac{1}{(4\pi t)^{\frac{d}{2}}} e^{-m^2t}  \sum_{p_0}  \Bigg[\tilde{b}_0 e^{Q^2t} -  \tilde{b}_1 t e^{Q^2t}  +  \tilde{B}_1 t^2 e^{Q^2t}
	+    \tilde{B}_2 t^2 Qe^{Q^2t}  \nonumber \\
	& &    - \Big[ \tilde{b}_3^{0} t^3\;e^{Q^2t} + \tilde{b}_{31}^{T} t^3\; e^{Q^2t} + \tilde{b}_{32}^{T}\;t^3\; Q\; e^{Q^2t} 
	+  \tilde{b}_{33}^{T}\;t^3\;Q^2\; e^{Q^2t} +  \tilde{b}_{34}^{T}\;t^3\;Q^3\; e^{Q^2t} \Big] \Bigg] \nn \\
	&=&  \frac{1}{(4\pi t)^{\frac{d+1}{2}}} e^{-m^2t} \Bigg[ \tilde{b}_0 \varphi_0 - \tilde{b}_1 t \varphi_0 +  \tilde{B}_1 t^2 \varphi_0 +  \tilde{B}_2 t^{3/2} \varphi_1  \nonumber \\
	& &    - \Big[  \tilde{b}_3^{0} t^3\;\varphi_0 +  \tilde{b}_{31}^{T} t^3\; \varphi_0 + \tilde{b}_{32}^{T}\;t^{5/2}\; \varphi_1 
	+  \tilde{b}_{33}^{T}\;t^2\; \varphi_2 +  \tilde{b}_{34}^{T}\;t^{3/2}\; \varphi_3 \Big] \Bigg].
\end{eqnarray}

Here, the  thermal HKCs (THKCs) are computed as (see appendix~\ref{HKC}),
\begin{eqnarray} \label{eq:thkc1}
\tilde{b}_0 &=& 1; \tilde{b}_1 = U; \nn \\
    \tilde{b}_2  & = & \tilde{b}_2^0 + \tilde{b}_{21}^T + Q \tilde{b}_{22}^T. \nonumber \\
\tilde{b}_3 	& =& (3!)\;\Big[\tilde{b}_3^{0} + \tilde{b}_{31}^{T} + Q\;\tilde{b}_{32}^{T} +  Q^2\;\tilde{b}_{33}^{T} + Q^3\;\tilde{b}_{34}^{T}\Big].
\end{eqnarray}
We identify the zero temperature HKC $\tilde{b}_2^0=[U^2 -\frac{1}{3} U_{ii}  + \frac{1}{6} (G_{ij})^2]$, and the finite temperature effects are captured through $\tilde{b}_{21}^T= \frac{1}{3} [ 2 E_i^2 + E_{ii0} - 3 U_{00}]$, $\tilde{b}_{22}^T= - \frac{2}{3} Q [E_{ii} -3 U_0]$. To proceed further let us define $\frac {\tilde{b}_2}{2!}= \tilde{B}_{1}+ Q \tilde{B}_{2}$ where $2 \tilde{B}_{1}= \tilde{b}_2^0 + \tilde{b}_{21}^T$, and $2 \tilde{B}_{2}= \tilde{b}_{22}^T$.
We also define,
\bea\label{eq:thkc2}
\tilde{b}_3^{0} &=& \frac{1}{3!} \Big[U^3  + \frac{1}{10} (J_i)^2 - \frac{1}{15} G_{ij} G_{jk} G_{ki} + \frac{1}{2} U (G_{ij})^2 + \frac{1}{2} (U_i)^2  + \frac{1}{10} U_{iijj} -\frac{1}{30} (J_{i})_{jji} \Big]; \nonumber  \\
 \tilde{b}_{31}^{T}  &=& \frac{1}{3!} \Big[  -  U_{0000} -2(U_0)^2 -UU_{00} +2U_{00} U +\frac{1}{2} [2 E_{i00} E_i + E_{i0}^2 - 2U_{i0} E_{i}  - U_i E_{i0} - E_{i0} U_i ] \nn \\
 &+& \frac{1}{10} [2 E_{ii} E_{jj} + 4 E_{ij} E_{ij} + 4 E_j E_{iij} + 2 E_i E_{ijj} +2 E_{ijj} E_i +E_{ii0jj}]  \Big] \nn \\
 \tilde{b}_{32}^{T} &=& \frac{1}{3!} \Big[  - 4 U_{000} +  2UU_0  - 4U_0 U
 + \frac{1}{2} [ 2  U_i E_{i} + 2 E_{i} U_i -6  E_{i0} E_i -2  E_i E_{i0} ] + \frac{1}{10} [ - 2  E_{iijj}] \Big]; \nn \\
 \tilde{b}_{33}^{T} &=&  \frac{1}{3!} \Big[  4 U_{00} - U_{ii}  + 2  E_i^2 +  E_{ii0}
  +  2 E_i^2 \Big]; \;\;  \tilde{b}_{34}^{T} =  \frac{1}{3!} \Big[ - 2 E_{ii} \Big].
\eea

Now, we are ready to write down the effective Lagrangian in terms of THKCs as
\begin{eqnarray}\label{eq:eft_lag_finite_temp}
	\mathcal{L}_{eff}&=& c_s\;\; \text{tr}\; \int_0^{\infty} \frac{dt}{t} K  \nonumber  \\
	& =& c_s\;\; \text{tr}\; \int_0^{\infty} \frac{dt}{t}\; \frac{\mu^{2\epsilon}e^{-m^2t}}{(4\pi t)^{\frac{d+1}{2}}} \; \Bigg[ \tilde{b}_0 \phi_0 - \tilde{b}_1 t \phi_0 +  \tilde{B}_1 t^2 \phi_0 +  \tilde{B}_2 t^{3/2} \phi_1 \nonumber \\
	& &    - \Big[  \tilde{b}_3^{0} t^3\;\phi_0 +  \tilde{b}_{31}^{T} t^3\; \phi_0 + \tilde{b}_{32}^{T}\;t^{5/2}\; \phi_1 
	+  \tilde{b}_{33}^{T}\;t^2\; \phi_2 +  \tilde{b}_{34}^{T}\;t^{3/2}\; \phi_3 \Big] \Bigg] \nonumber \\
	& =& c_s\;\; \text{tr}\; \Bigg[  \tilde{b}_0 \mathbb{I}[0;0] - \tilde{b}_1 \mathbb{I}[0;1]  + \tilde{B}_1 \mathbb{I}[0;2] + \tilde{B}_2 \mathbb{I}[1;3/2] \nonumber \\
	& & - \Big[  \tilde{b}_3^{0} \mathbb{I}[0;3] +  \tilde{b}_{31}^{T} \mathbb{I}[0;3] + \tilde{b}_{32}^{T}\;\mathbb{I}[1;5/2]
	+  \tilde{b}_{33}^{T}\;\mathbb{I}[2;2] +  \tilde{b}_{34}^{T}\;\mathbb{I}[3;3/2]  \Big] \Bigg]    .  
\end{eqnarray}
Here, we define the following integral
\begin{eqnarray}
	\mathbb{I} [k;l] &=& \int_0^{\infty} \frac{dt}{t}\; \frac{\mu^{2\epsilon}e^{-m^2t}}{(4\pi t)^{\frac{d+1}{2}}}\; t^l \;\phi_k (\Omega; t/\beta^2) \nonumber \\
	& =& \int_0^{\infty} \frac{dt}{t}\; \frac{\mu^{2\epsilon}e^{-m^2t}}{(4\pi t)^{\frac{d}{2}}}\; t^l\; \frac{1}{\beta} \; \sum_{p_0=\frac{2\pi n }{\beta}} t^{k/2} Q^k e^{Q^2 t}.
\end{eqnarray}
Recall that, here $Q= D_0 + \iu  p_0 \equiv \iu \frac{2n\pi}{\beta}-\frac{1}{\beta} Log(\Omega)= \frac{2\pi  \iu }{\beta} [n+ \frac{ \iu }{2\pi} Log(\Omega)]$. Define, $\tilde{n} = \frac{ \iu }{2\pi} Log(\Omega) \in \mathbb{R}$, and $\tilde{n} \notin \mathbb{Z}$ where  
$n \in \mathbb{Z}$.
Thus, the previous integral can be written in the following form
\begin{eqnarray} \label{Ikl}
	\mathbb{I}[k;l] &=& \int_0^{\infty} \frac{dt}{t}\; \frac{\mu^{2\epsilon}e^{-m^2t}}{(4\pi t)^{\frac{d}{2}}}\; t^l\; \frac{1}{\beta} \; \sum_{n} t^{k/2} \Big[\frac{2\pi  \iu }{\beta}\Big]^k [n+\tilde{n}]^k e^{Q^2 t} \nonumber \\
	&=&  \sum_{n}  \int_0^{\infty} \frac{dt}{t}\; \frac{\mu^{2\epsilon}}{\beta} 
	\Big[\frac{2\pi  \iu }{\beta}\Big]^k [n+\tilde{n}]^k t^{(2l+n-d)/2} e^{-m^2t} \frac{e^{Q^2 t}}{(4\pi)^{d/2}} \nonumber \\
	& =& \sum_{n}  \Big[\frac{2\pi  \iu }{\beta}\Big]^k \frac{\mu^{2\epsilon}}{\beta (4\pi)^{d/2}} 
	[n+\tilde{n}]^k \int_0^{\infty}\; dt\; t^{(2l+k-d-2)/2}  e^{-m^2t} e^{-(\frac{2\pi}{\beta})^2 (n+\tilde{n})^2 t}.
\end{eqnarray}
Define, $N= m^2+ (\frac{2\pi}{\beta})^2 (n+\tilde{n})^2$, and $R= (2l+k-d-2)/2$, then we have the following integral 
\begin{eqnarray}
	\mathbb{I} &=&  \int_0^{\infty} dt\; t^{R} e^{-Nt} 	= \frac{1}{|N|^{R+1}} \Gamma (R+1)  \nonumber \\
	&=& \Big[ m^2+ \Big(\frac{2\pi}{\beta}\Big)^2 (n+\tilde{n})^2\Big]^{-\frac{2l+k-d}{2}}  
	\Gamma \Big(\frac{2l+k-d}{2}\Big),
\end{eqnarray}
which leads us in (\ref{Ikl}) the form for the different $\mathbb{I}(k,l)$ as in terms of a summation over the Matsubara frequency,
\bea
\mathbb{I} [k;l] &=&\sum_{n}  \Big[\frac{2\pi  \iu }{\beta}\Big]^k \frac{\mu^{2\epsilon}}{\beta (4\pi)^{d/2}} 
	[n+\tilde{n}]^k \; \Big[ m^2+ \Big(\frac{2\pi}{\beta}\Big)^2 (n+\tilde{n})^2\Big]^{-\frac{2l+k-d}{2}}  
	\Gamma \Big(\frac{2l+k-d}{2}\Big).\nn
\eea

We  perform the sum over the Matsubara frequency as follows,
\begin{eqnarray}
	\sum_{n}  
	[n+\tilde{n}]^k   \Big[ m^2+ \Big(\frac{2\pi}{\beta}\Big)^2 (n+\tilde{n})^2\Big]^{-\frac{2l+k-d}{2}}
	& = & \sum_{n}  \Big(\frac{2\pi}{\beta}\Big)^{(d-2l-k)}  [n+\tilde{n}]^k   \Big[ (\frac{\beta m }{2 \pi})^2+  (n+\tilde{n})^2\Big]^{-\frac{2l+k-d}{2}} \nonumber \\
	& = & \sum_{n}  \Big(\frac{2\pi}{\beta}\Big)^{(d-2l-k)} \frac{[n+\tilde{n}]^k}{|(n +\tilde{n})+ I\;m_{\beta}|^{k+2l-d}} \nonumber \\
	& = &  \Big(\frac{2\pi}{\beta}\Big)^{(d-2l-k)} \frac{(-1)^k}{\Pi_{j=1}^k (2l-d+j)} \sum_{n} \frac{d^k\;G(x)}{d x^k}\Bigg|_{x=1} .\nonumber
\end{eqnarray}
Here, we define $m_{\beta}= \frac{ m \beta}{2 \pi}$, and $\tilde{n}=\frac{ \iu }{2\pi} ln(\Omega)$, $(n +\tilde{n})^2+m_{\beta}^2=|(n +\tilde{n})+  \iu \;m_{\beta}|^2 $ and we introduced the  function $G(k;x)=\frac{1}{|(n +\tilde{n})x+  \iu \;m_{\beta}|^{2l-d}}$ where $x \in \mathbb{R}^{+}$. The summation of this function over Matsubara frequencies can
then be written in terms of  Epstein-functions.  We can write 
\begin{eqnarray}
	\sum_{n=-\infty}^{\infty} G(k;x) & = &  \sum_{n=-\infty}^{\infty} \frac{1}{|(n +\tilde{n})x+  \iu \;m_{\beta}|^{2l-d}} \nonumber \\
	&=& \frac{1}{|x|^{(2l-d)}}  \sum_{n=-\infty}^{\infty} \frac{1}{\Big[(n +\tilde{n})^2+ (\;m_{\beta}/x)^2\Big]^{(2l-d)/2}} \nonumber \\
	& = &   \frac{1}{|x|^{(2l-d)}}  \Big[	  C^{1-2s} \sqrt{\frac{\pi}{a_1}} \frac{\Gamma(s-1/2)}{\Gamma(s)}  \nonumber \\
	& + & \frac{4 \pi^s}{\Gamma(s)}  a_1^{-s/2-1/4} C^{-s+1/2} \sum_{k=1}^{\infty} \cos(2\pi b_1 k)\; k^{s-1/2} \mathbb{K}_{s-1/2}(2\pi k\; C) \Big],\nn\\
	& = &  \frac{1}{|x|^{(2l-d)}} \Bigg[   	\mathbb{E}^{(m_\beta/x)} \Big(\frac{2l-d}{2};\; 1;\; \tilde{n} \Big)      \Bigg],
\end{eqnarray}
where $C=m_\beta/x, a_1=1, b_1=\tilde{n}, s=(2l-d)/2$, and we define \cite{Elizalde:1994gf}
\begin{eqnarray}\label{Epstein-1}
		\mathbb{E}^{(m_\beta/x)} \Big(\frac{2l-d}{2};\; 1;\; \tilde{n} \Big)& =&  \Big[	  C^{1-2s} \sqrt{\frac{\pi}{a_1}} \frac{\Gamma(s-1/2)}{\Gamma(s)}  \nonumber \\
		& + & \frac{4 \pi^s}{\Gamma(s)}  a_1^{-s/2-1/4} C^{-s+1/2} \sum_{k=1}^{\infty} \cos(2\pi b_1 k)\; k^{s-1/2} \mathbb{K}_{s-1/2}(2\pi k\; C) \Big]. \nn
\end{eqnarray}

Therefore the sum over the Matsubara frequency can be expressed as 
\begin{eqnarray}
	&& \sum_{n}    [n+\tilde{n}]^k   \Big[ m^2+ \Big(\frac{2\pi}{\beta}\Big)^2 (n+\tilde{n})^2\Big]^{-\frac{2l+k-d}{2}}  \\
	& = &\Big(\frac{2\pi}{\beta}\Big)^{(d-2l-k)} \frac{(-1)^k}{\Pi_{j=1}^k (2l-d+j)}  \frac{d^k}{d x^k} \Big[ \frac{1}{|x|^{(2l-d)}}\; 
	\Bigg[   	\mathbb{E}^{(m_\beta/x)} \Big(\frac{2l-d}{2};\; 1;\; \tilde{n} \Big)      \Bigg] \Bigg|_{x=1} \Big] .\nonumber
\end{eqnarray}

Now, after performing the sum the final and compact form of the integral is given as
\begin{eqnarray}\label{eq:bosonic_integrals_1}
	\mathbb{I}[k;l] &=&
	\Big[\frac{2\pi  \iu }{\beta}\Big]^k \frac{\mu^{2\epsilon}}{\beta (4\pi)^{d/2}} 
	\Big(\frac{2\pi}{\beta}\Big)^{(d-2l-k)} \frac{(-1)^k}{\Pi_{j=1}^k (2l-d+j)} \Gamma (\frac{2l+k-d}{2})
	\nonumber \\ & & \frac{d^k}{d x^k} \Big[ \frac{1}{|x|^{(2l-d)}}\; 
	\Bigg[   	\mathbb{E}^{(m_\beta/x)} \Big(\frac{2l-d}{2};\; 1;\; \tilde{n} \Big)
     \Bigg] \Big] \Bigg|_{x=1} .   
\end{eqnarray}
Here, we must set $d=3-2\epsilon$ to recover the thermally corrected effective action for a four-dimensional theory. 


\subsection{Effective Action: Without Polyakov Loops}\label{EffNPL}

We use the generalised  Chowla-Selberg formula \cite{CSF} derived in \cite{Elizalde-1, Elizalde-2}, in the limit  in which Polyakov loop contribution vanishes, i.e., $b_1=\tilde{n}=0$, as $\zeta (-2m; 0)=0$ for $m\in Z^+$:
\begin{eqnarray}
	\mathbb{CS}(r;y)=   \sum_{k=1}^{\infty} \frac{1}{(k^2+r^2)^y} &=& -\frac{r^{-2y}}{2} + \frac{\sqrt{\pi}\; \Gamma(y-1/2)}{2 \Gamma(y)} r^{-2y+1} \nonumber \\
	& +& \frac{2 \pi^y r^{-y+1/2}}{\Gamma(y)} \sum_{k=1}^{\infty} k^{y-1/2} 
	\; \mathbb{K}_{y-1/2} (2\pi k\;r),
\end{eqnarray}
where $r \in \mathbb{R} \notin \mathbb{Z}$. 

In the limit when we set the Polyakov loop contribution to zero, i.e.,  $\tilde{n} \to 0\equiv \Omega \to 1$, we find, 
\bea \label{I00-1}
\mathbb{I}[0;0] &=&   \frac{\mu^{2\epsilon}}{\beta (4\pi)^{d/2}} \Big(\frac{2\pi}{\beta}\Big)^{d}\;\;
\Gamma (\frac{-d}{2})\Bigg[ ( \;m_{\beta})^{+d} 
+ 2\; \mathbb{CS} \Big(m_{\beta};\; \frac{-d}{2} \Big) \Bigg]  \nonumber \\
& =& 2\; \Bigg\{ \frac{m^2 }{2\pi^2 \beta^2} \sum_{n=1}^{\infty} \frac{1}{n^2} 
\mathbb{K}_{-2} (n m \beta) + \frac{m^4}{32\pi^2} \Gamma(-2+\epsilon) \Big(\frac{2\pi \mu^2}{m^2}\Big)^{\epsilon} \Bigg\} \nn \\
&=&  \; \frac{m^4}{32\pi^2} \Big(\ln(\frac{4\pi \mu^2}{m^2})+\frac{3}{2}-\gamma_E\Big) +
\frac{m^2 }{ \pi^2 \beta^2} \sum_{n=1}^{\infty} \frac{1}{n^2} \mathbb{K}_{-2} (n m \beta).
\eea

We can evaluate the expressions for small and large  $ m \beta $ as follows. For high temperatures ($m\beta < \pi/2$) we sum the Bessel series as shown in Appendix~\ref{app:sum_bessel} and obtain
\bea\label{I00s}
\mathbb{I}[0;0]&=&  \;  \frac{m^4}{32 \pi^2} \Big(\ln(\frac{4\pi \mu^2}{m^2})+\frac{3}{2}-\gamma_E\Big)+  \mathbb{S}[0;0]   \nn\\
&=&   \;   \frac{m^4}{32 \pi^2} \Big(\ln(\frac{4\pi \mu^2}{m^2})+\frac{3}{2}-\gamma_E\Big)    \nn \\
& + &  \frac{m^3}{6 \pi \beta} \; +\; \frac{\pi^2}{45\; \beta^4} - \frac{ m^2}{12\; \beta^2} +\frac{m^4}{ (4\pi)^2}
\Big( ln\Big[\frac{m\beta e^{\gamma_E}}{4\pi} \Big] -\frac{3}{4}\Big )\nn\\
& -&\frac{2\zeta(3) \; m^6 \; \beta^2}{3 \; (4\pi)^4} + +\frac{ \zeta (5) m^8 \beta ^4 }{ (4 \pi) ^6}+\cdots. 
\eea
This result agrees with the expression of pressure of bosons at high temperatures \cite{Dolan:1973qd}
\bea
p=c_s \mathbb{I}[0;0]
&=&      \frac{m^3}{12 \pi \beta} \; +\; \frac{\pi^2}{90\; \beta^4} - \frac{ m^2}{24\; \beta^2} +\frac{m^4}{2\; (4\pi)^2}
\Big( ln\Big[\frac{m\beta e^\gamma}{4\pi} \Big] -\frac{3}{4}\Big )\nn\\
& -&\frac{\zeta(3) \; m^6 \; \beta^2}{3 \; (4\pi)^4} +\frac{ \zeta (5) m^8 \beta ^4 }{2 (4 \pi) ^6}+\cdots .\nn
\eea

For low temperatures $m \beta > \pi/2$ we can expand the Bessel function in (\ref{I00-1}) the large argument limit,
\be\label{Knuz}
 \mathbb{K}_{\nu} (z) = \Big(\frac{\pi}{2z}\Big)^{1/2} e^{-z} \sum_{k=0}^{\infty} \frac{ (\frac{1}{2}-\nu)_k (\frac{1}{2}+\nu)_k}{(-2)^k\,\, k!} \frac{1}{z^k}\;\;,
\ee
where we have introduced the Pochhammer symbol $(~)_k$ defined as $(x)_k=\Gamma[x+k]/\Gamma[x]$. We see that due to the exponential suppression, for $m\beta > \pi/2$,  the $k=1$ term in the Bessel sum in (\ref{I00-1}) dominates. Retaining just the $k=1$ term in (\ref{I00-1}) and evaluating the leading terms in  the Bessel function expansion  (\ref{Knuz}), we obtain the  expression for the boson pressure at the low temperature given by, 
\bea
{\cal L}=c_s\,\mathbb{I}[0;0]= T^4 \Big(\frac{m}{2 \pi T}\Big)^{3/2} e^{-m/T}\Big[1+\frac{15}{8} \frac{T}{m} &+&  \frac{108}{128} \Big(\frac{T}{m}\Big)^2 -  \frac{315}{1024} \Big(\frac{T}{m}\Big)^3 \nn\\
\nn\\
&+&   \frac{10395}{32768} \Big(\frac{T}{m}\Big)^4 +\cdots  \Big].
\eea

We can evaluate the other coefficients $\mathbb{I}[l;k]$ in the high and low $(m\beta)$ limits in the same way. The expressions for some other $\mathbb{I}[l;k]$'s in the $m \beta< \pi/2$ limit useful for the computation of phase transitions in effective theories up to dimension-6 are calculated in Appendix~\ref{app:sum_bessel} and are listed below
\bea\label{Ikl-list}
\mathbb{I}[0;1] &=&  \frac{\mu^{2\epsilon}}{\beta (4\pi)^{d/2}} \Big(\frac{2\pi}{\beta}\Big)^{d-2}\;\;
\Gamma (\frac{2-d}{2}) \Bigg[ (\;m_{\beta})^{-2+d} 
+ 2\; \mathbb{CS} \Big(m_{\beta};\; \frac{2-d}{2} \Big) \Bigg] \nonumber  \\
&=& \frac{m^2}{8\pi^2} \left( \ln(\frac{4\pi m^2}{\mu^2})-1+\gamma_E \right)+\frac{m}{2\pi^2 \beta} \sum_{n=1}^{\infty} \frac{1}{n} 
\mathbb{K}_{-1} (n m \beta) \nn\\
&=& \Big\{ \frac{m^2}{8\pi^2} \left( \ln(\frac{4\pi m^2}{\mu^2})-1+\gamma_E \right)+ \mathbb{S}[0;1] \Big \} \nn\\
&=& \frac{m^2}{8\pi^2} \left( \ln(\frac{4\pi m^2}{\mu^2})-1+\gamma_E \right)\nn\\
&-&\frac{m}{4\pi \beta} + \frac{1}{12\beta^2} -\frac{m^2}{8\pi^2} \Big\{ ln\Big[ \frac{m\beta e^\gamma}{4\pi}\Big] -\frac{1}{2}\Big \} +\frac{2m^4 \beta^2}{(4\pi)^4} \zeta(3) +\cdots
\eea

\bea
\mathbb{I}[0;2]  &=&  \frac{\mu^{2\epsilon}}{\beta (4\pi)^{d/2}} \Big(\frac{2\pi}{\beta}\Big)^{d-4}\;\;
\Gamma (\frac{4-d}{2}) \Bigg[ (\;m_{\beta})^{-4+d} 
+ 2\; \mathbb{CS} \Big(m_{\beta};\; \frac{4-d}{2} \Big) \Bigg]  \nonumber \\
 &=&\frac{1}{16 \pi^2}\left(\ln(\frac{4 \pi \mu^2}{m^2})-\gamma_E\right)+ \frac{1}{4 \pi^2  }   \sum_{n=1}^{\infty}  \mathbb{K}_{0} (n m \beta) \nn\\
&=& \frac{1}{16 \pi^2}\left(\ln(\frac{4 \pi \mu^2}{m^2})-\gamma_E\right)+ \mathbb{S}[0;2]\nn\\
&=& \frac{1}{16 \pi^2}\left(\ln(\frac{4 \pi \mu^2}{m^2})-\gamma_E\right)+       \frac{1}{ 8\pi m \beta} + \frac{1}{4\pi^2} \Big\{  \frac{ 1}{2}  ln\Big[ \frac{m\beta\; e^\gamma}{4\pi}\Big] - \frac{\zeta(3) \Gamma(3)}{2 (2\pi)^2} \Big[ \frac{m\beta}{2} \Big]^2 \nn \\
&+&  \frac{\zeta(5) \Gamma(5)}{2 (2\pi)^4} \Big[ \frac{m\beta}{2} \Big]^4 + \cdots   \Big\} .
\eea

\bea
\mathbb{I}[0;3] &=&  \frac{\mu^{2\epsilon}}{\beta (4\pi)^{d/2}} \Big(\frac{2\pi}{\beta}\Big)^{d-6}\;\;
\Gamma (\frac{6-d}{2}) \Bigg[ ( \;m_{\beta})^{-6+d} 
+ 2\; \mathbb{CS} \Big(m_{\beta};\; \frac{6-d}{2} \Big) \Bigg]  \nonumber \\
&=& \frac{1}{16 \pi^2 m^2}+ \frac{1}{8 \pi^2 m T }   \sum_{n=1}^{\infty} n\, \mathbb{K}_{1} (n m \beta)\nn\\
&=&\frac{1}{16 \pi^2 m^2}+ \mathbb{S}[0;3]\nn\\
&=& \frac{1}{16 \pi^2 m^2}+\frac{1}{8\pi m^3 \beta} + \frac{ \beta^2}{2 (2\pi)^2} \zeta(3) -\frac{12m^2 \beta^4}{(4\pi)^6 } \zeta(5)+\cdots
\eea

 Employing the Heat-Kernel coefficients listed in (\ref{eq:thkc1}) and (\ref{eq:thkc2}) with the thermal factors $\mathbb{I}(k,l)$,  the complete effective action is given as 
 \begin{eqnarray}\label{eq:eft_lag_finite_temp_boson}
 	\mathcal{L}_{eff}&=& c_s\;\;  \Bigg[ \text{tr}[\tilde{b}_0]\;\; \mathbb{I}[0;0] - \text{tr}\;[\tilde{b}_1] \;\; \mathbb{I}[0;1] +  \text{tr}\;[\tilde{B}_1]\;\; \mathbb{I}[0;2] +  \text{tr}\;[\tilde{B}_2] \;\; \mathbb{I}[1;3/2] \nonumber \\
 	& & - \text{tr} \Big[  \tilde{b}_3^{0} \mathbb{I}[0;3] +  \tilde{b}_{31}^{T} \mathbb{I}[0;3] + \tilde{b}_{32}^{T}\;\mathbb{I}[1;5/2]
 	+  \tilde{b}_{33}^{T}\;\mathbb{I}[2;2] \Big] \Bigg]  \\
 	&=& c_s\;\; \text{tr} \Bigg[ \mathbb{I}[0;0] - U \;\;\mathbb{I}[0;1] + \frac{1}{2} \Big[\Big(U^2 -\frac{1}{3} U_{ii}  + \frac{1}{6} (G_{ij})^2 \Big) + \frac{1}{3} \Big( 2 E_i^2 + E_{ii0} - 3 U_{00}\Big) \Big] \;\; \mathbb{I}[0;2] \nonumber \\
 	& &  - \frac{1}{3}  \Big(E_{ii} -3 U_0\Big) \;\; \mathbb{I}[1;3/2] 
 	- \Big[  \tilde{b}_3^{0} \mathbb{I}[0;3] +  \tilde{b}_{31}^{T} \mathbb{I}[0;3] + \tilde{b}_{32}^{T}\;\mathbb{I}[1;5/2]
 	+  \tilde{b}_{33}^{T}\;\mathbb{I}[2;2] +  \tilde{b}_{34}^{T}\;\mathbb{I}[3;3/2] \Big] \Bigg]. \nn
 \end{eqnarray}    

 
 \subsection{Effective Action: With Polyakov Loops}
When the scalars and fermions in the loop or the heat bath carry gauged charges, the thermal factors of the HKC are from  (\ref{Epstein-1})  and (\ref{eq:bosonic_integrals_1}), given by 
\begin{eqnarray}\label{non-pert-PL}
\mathbb{I}[0;0] 
&=&   \frac{m^4}{32\pi^2} \Big(\ln(\frac{4\pi \mu^2}{m^2})+\frac{3}{2}-\gamma_E\Big) +
\frac{m^2 }{ \pi^2 \beta^2} \sum_{n=1}^{\infty} \frac{1}{n^2} \cos(2\pi n\tilde{n}) \mathbb{K}_{-2} (n m \beta),   \nn\\
\mathbb{I}[0;1]  &=& \frac{m^2}{8\pi^2} \left( \ln(\frac{4\pi m^2}{\mu^2})-1+\gamma_E \right)+\frac{m}{2\pi^2 \beta} \sum_{n=1}^{\infty} \frac{1}{n} \cos(2\pi n\tilde{n})
\mathbb{K}_{-1} (n m \beta), \nn\\
\mathbb{I}[0;2]  &=&\frac{1}{16 \pi^2}\left(\ln(\frac{4 \pi \mu^2}{m^2})-\gamma_E\right)+ \frac{1}{4 \pi^2  }   \sum_{n=1}^{\infty}  \cos(2\pi n\tilde{n})  \mathbb{K}_{0} (n m \beta), \nn\\
\mathbb{I}[0;3]  &=& \frac{1}{16 \pi^2 m^2}+ \frac{1}{8 \pi^2 m T }   \sum_{n=1}^{\infty} n\,  \cos(2\pi n\tilde{n})  \mathbb{K}_{1} (n m \beta).
\end{eqnarray}
We evaluate the expressions for $\mathbb{I}_\Omega [0,0]$ with Polyakov loop contribution in for the case $m \beta < \pi/2$  and $\tilde n <1/(2 \pi)$ as follows.
In this limit, we can rewrite the sum as follows (see appendix for detailed evaluation)
\begin{eqnarray}\label{SOmega-1}
\mathbb{S}_\Omega [0,0] &=& \frac{m^2 }{ \pi^2 \beta^2} \sum_{n=1}^{\infty} \frac{1}{n^2} \cos(2\pi n\tilde{n}) \mathbb{K}_{-2} (n m \beta) \nn \\
&=& \frac{m^2 }{ \pi^2 \beta^2} \sum_{n=1}^{\infty} \sum_{q=0}^{\infty} \frac{1}{n^2} \frac{(2\pi n \tilde{n})^{2q} (-1)^q}{(2q)!} \mathbb{K}_{-2} (n m \beta) \nn \\
&=& \frac{m^2 }{ \pi^2 \beta^2} \sum_{q=0}^{\infty} \frac{1}{n^2} \frac{(2\pi \tilde{n})^{2q} (-1)^q}{(2q)!} 
\sum_{n=1}^{\infty} \frac{1}{n^{2-2q}} \mathbb{K}_{-2} (n m \beta) \nn \\
&=& \sum_{q=0}^{\infty}  \frac{(2\pi \tilde{n})^{2q} (-1)^q m^2}{(2q)! (\pi^2 \beta^2)} \Bigg\{ \frac{1}{4} \Big[\frac{(m\beta)}{2}\Big]^{1-2q} \Gamma(q+1/2) \Gamma(-q-3/2) + \frac{1}{2} \Big [ \frac{4 \zeta(4-2q)}{(m\beta)^2} - \zeta(2-2q) \Big ]  \nn \\
&+ & 
 \Big[ -\frac{m\beta}{2}\Big]^{2} \frac{1}{ 2!}
\Big \{  \zeta^{'}(-2q)  
  + \zeta(-2q) \Big[  \frac{1}{2} \psi(1) + \frac{1}{2} \psi(3) -ln(m\beta/2) \Big] \Big \}  \nn \\
& + & \sum_{r=1}^{\infty} \Big[ -\frac{m\beta}{2}\Big]^{2r+2} \frac{1}{r! (r+2)!}
\Big \{ \frac{1}{2} \Big [  -\frac{1}{(2\pi)^2}\Big ]^{r+q} \zeta(2r+2q+1) \Gamma(2r+2q+1) \Bigg \}.
\end{eqnarray}
Evaluating the terms up to $(\beta m )^6$ in (\ref{SOmega-1}), we obtain the expression for the pressure in the high temperature $m \beta <\pi/2$ and as a perturbative expansion in the Polyakov loop factor $\tilde n< 1/2\pi$ given by
 \bea\label{S00PL}
\mathbb{S}_\Omega [0,0]&=&-\frac{m^4}{32 \pi^2 }\left(\log \left(\frac{\beta  m }{4 \pi}\right)
-\frac{3}{4}+  {\tilde n}^2 \zeta (3) \right)- \frac{\pi ^2 \left(1-30 {\tilde n}^2\right)}{45 \beta ^4} \nn\\
&+&
\frac{\left(1+6 {\tilde n}^2\right) m^2}{12 \beta ^2}-\frac{4 \pi  {\tilde n}^2 m}{15 \beta ^3}-\frac{m^3}{6 \pi  \beta }+\frac{ \beta ^2 m^6 \left(6 {\tilde n}^2 \zeta (5)+\zeta (3)\right)}{384 \pi ^4}.
\eea

 For low-temperature applications, we compute   
 \bea
 {\cal L}_\Omega =c_s \mathbb{S}_\Omega[0;0] 
=  \frac{m^2 }{2 \pi^2 \beta^2} \sum_{n=1}^{\infty} \frac{1}{n^2} \cos(2\pi n\tilde{n}) \mathbb{K}_{-2} (n m \beta),
\eea
in the  $m \beta > \pi/2$ limit. Using (\ref{Knuz}), we see that  the $n=1$ dominates exponentially, and we obtain the expression for the boson pressure with Polyakov loop contribution given by 
\bea
{\cal L}_\Omega= T^4 \Big(\frac{m}{2 \pi T}\Big)^{3/2} e^{-m/T}\,\cos(2\pi \tilde{n})\,\Big[1+\frac{15}{8} \frac{T}{m} &+&  \frac{108}{128} \Big(\frac{T}{m}\Big)^2 -  \frac{315}{1024} \Big(\frac{T}{m}\Big)^3 \nn\\
\nn\\
&+&   \frac{10395}{32768} \Big(\frac{T}{m}\Big)^4 +\cdots  \Big].
\eea

 
  \section{Integrating out  Fermions} \label{sec:fermion}
  
In the case of fermions  \cite{Chakrabortty:2023yke}, the Lagrangian is given as 
\begin{equation}
	\mathcal{L}_f = \overline{\Psi} (P - m_f -\Sigma) \Psi,
\end{equation}
where $\Sigma = S+  \iu  R \gamma_5$ with $S,R$ are the scalar fields. The operator $(P - m_f -\Sigma)$ is weak-elliptic operator. After performing bosonization, we can rewrite this operator as a strong elliptic operator and compute the effective action following the similar HK method adopted for bosons. Here, the functional form of the HKCs remains unaltered after the following redefinition \cite{Chakrabortty:2023yke}
\begin{eqnarray}\label{eq:ferm_hkc}
	P_\mu & \to & P_\mu -  \iu \gamma^5 \gamma_\mu R;\\
	U &\to & U_f= 2m_f \Sigma -\frac{1}{2} \sigma_{\mu \nu} G^{\mu \nu} + S^2 + 3R^2 - (PS) +  \iu \gamma^5 (RS+SR); \\
	G_{\mu \nu} & \to & G_{\mu \nu} +  \iu  \gamma^5 [\gamma_\mu P_\nu R - \gamma_\nu P_\mu R] + 2 \sigma_{\mu \nu} R^2.
\end{eqnarray}
Here, we consider the spin-0 fields  $(R, S)$, and the gauge fields are background ones. Then, we can write down the zero temperature HKCs for the fermion case as 
\begin{equation}
	b_0=1, b_1= U_f, b_2 = U_f^2 + \frac{1}{6} G_{\mu \nu}^2 -\frac{1}{3} {U_f}_{\mu\mu}.
\end{equation}
Effective action after integrating out heavy fermion, we find a very similar effective action, given in (\ref{eq:eft_lag_finite_temp}).
\begin{eqnarray}\label{eq:eft_lag_finite_temp_ferm}
	\mathcal{L}_{eff} &=&  c_f\;\; \text{tr}\; \Bigg[  \tilde{b}_0 \mathbb{I}_F[0;0] - \tilde{b}_1 \mathbb{I}_F[0;1]  + \tilde{B}_1 \mathbb{I}_F[0;2] + \tilde{B}_2 \mathbb{I}_F[0;3/2] \nonumber \\
	& & - \Big[  \tilde{b}_3^{0} \mathbb{I}_F[0;3] +  \tilde{b}_{31}^{T} \mathbb{I}_F[0;3] + \tilde{b}_{32}^{T}\;\mathbb{I}_F[1;5/2]
	+  \tilde{b}_{33}^{T}\;\mathbb{I}_F[2;2] \Big] \Bigg] .     
\end{eqnarray}
Note that the THKCs $(\tilde{b}_i, \tilde{b}_{ij}, \tilde{B}_i)$, for fermions, are of similar forms as for bosonic case, given in previous section, incorporating the redefinition given in Eq.~\ref{eq:ferm_hkc}  with an additional trace over gamma-matrices as the HKCs are matrices in Clifford space.

 Here, the $\phi_n$'s are the thermal wave function, which is anti-periodic for fermions. We can still use the previous results, i.e., bosonic case, with a replacement  $\Omega \to - \Omega$, and that implies that in case of fermions, $\tilde{n}_f = \frac{ \iu }{2\pi} log(-\Omega) \equiv (\frac{1}{2}+ \tilde{n}_b) \in \mathbb{R}$, and $\tilde{n} \notin \mathbb{Z}$. Thus, $\tilde{n}_f= \tilde{n}_b+1/2$, where $\tilde{n}_b$ is the $\tilde{n}$ in the bosonic case, see the previous section. 

In the case of fermion 
\begin{eqnarray}
	\mathbb{I}_F[0;0] (\beta) &=& \sum_{n} \frac{\mu^{2\epsilon}}{\beta (4\pi)^{d/2}} 
	\Big[ m^2+ \Big(\frac{2\pi}{\beta}\Big)^2 ((n+1/2)+\tilde{n})^2\Big]^{\frac{d}{2}}
	\Gamma (\frac{-d}{2}) \nonumber \\
	&=& 2  \mathbb{I}_B[0;0] (2\beta) -  \mathbb{I}_B[0;0] (\beta), \nonumber \\
	\mathbb{I}_F[0;1] (\beta) &=& \sum_{n}  \frac{\mu^{2\epsilon}}{\beta (4\pi)^{d/2}}  
	\Big[ m^2+ \Big(\frac{2\pi}{\beta}\Big)^2 ((n+1/2)+\tilde{n})^2\Big]^{\frac{d-2}{2}}
	\Gamma (\frac{2-d}{2}) \nonumber \\
	&=& 2 \mathbb{I}_B[0;1] (2\beta) -   \mathbb{I}_B[0;1] (\beta) .
\end{eqnarray}
 Note that zero-mode contributions from  $\mathbb{I}_B[0;0] (2\beta)$ and $\mathbb{I}_B[0;0] (\beta)$ exactly   cancel each other. 
In general, we can write $	\mathbb{I}_F[k;l] = 2 \mathbb{I}_B[k;l] (2\beta) -   \mathbb{I}_B[k;l] (\beta)  $ $\forall \{k;l\}$, see appendix for the general proof. For the case of zero Polyakov loops, using the  bosonic integral  $\mathbb{I}_B[k;l] $  given in (\ref{eq:bosonic_integrals_1}) and listed in (\ref{Ikl-list}), we find:
\begin{eqnarray}\label{fermionPL}
	\mathbb{I}_F[0;0] 	&=&      2\; \Bigg\{  \frac{m^4}{64 \pi^2} \Big(\ln(\frac{4\pi \mu^2}{m^2})+\frac{3}{2}-\gamma_E\Big)   
 +  \Big[  -\; \frac{7 \pi^2}{8 (90\; \beta^4)} + \frac{ m^2}{2(24\; \beta^2)} \nonumber \\
 & & +\frac{m^4}{2\; (4\pi)^2}
\Big\{ ln\Big[\frac{m\beta e^\gamma}{\pi} \Big] -\frac{3}{4} \Big \} - \frac{7 \zeta(3) \; m^6 \; \beta^2}{3 \; (4\pi)^4} +\cdots\Big]  \Bigg \} . \nn\\
\label{eq:fermionic_integrals_0}
\end{eqnarray} 

With Polyakov loops, we have the finite temperature contribution of fermions in the loop to the effective action, for $\tilde n <1/(2\pi)$, is  given by:
\bea\label{PLFL-1}
\mathbb{S}_{\Omega F}(0,0)&=&-\frac{7}{8}\frac{ \pi ^2 \left(1-30 {\tilde n}^2\right)}{90 \beta ^4}-\frac{7 \beta ^2 m^6 \left(6 {\tilde n}^2 \zeta (5)+\zeta (3)\right)}{3 (2 \pi )^4}\nn\\
&+&\frac{\left(6 {\tilde n}^2+1\right) m^2}{48 \beta ^2}-\frac{\pi  {\tilde n}^2 m}{10 \beta ^3}+\frac{m^4 }{64 \pi ^2} \left(ln \Big(\frac{\beta  m}{4\pi}\Big)-\frac{3}{4}+ \tilde n^2 \zeta(3) \right)+\cdots
\eea

If there are no background scalar fields but we have background gauge fields, then we have $U_f = -\frac{1}{2} \sigma_{\mu \nu} G^{\mu \nu} = -\frac{1}{2} \sigma^{\mu \nu} \sum_{a=1}^{N^{2}-1} G_{\mu \nu}^a T^a$ with $T^a$s are generators of background $SU(N)$ gauge theory. Thus, $tr[U_f]=0$ as $tr[T^a]=0$, and the effective action is given as
\begin{eqnarray}\label{EHL}
	\mathcal{L}_{eff}&=& c_f\;\; \text{tr}\; \int_0^{\infty} \frac{dt}{t} K  \nonumber  \\
	& =& c_f\;\; \text{tr}\; \int_0^{\infty} \frac{dt}{t}\; \frac{\mu^{2\epsilon}e^{-m^2t}}{(4\pi t)^{\frac{d+1}{2}}} \; \Bigg[ \tilde{b}_0 \phi_0 - \tilde{b}_1 t \phi_0 +  \tilde{B}_1 t^2 \phi_0 +  \tilde{B}_2 t^{3/2} \phi_1 \Bigg] \nonumber \\
	&=& c_f\;\; \text{tr} \Bigg[ \mathbb{I}_{F}[0;0] + \frac{1}{2} \Big[\frac{4}{3} \; [(G_{ij})^2+E_i^2-\frac{1}{3} {U_f}_{ii}] + \frac{1}{3} \Big( 2 E_i^2 + E_{ii0} - 3 U_{00}\Big) \Big] \;\; \mathbb{I}_{F}[0;2] \nonumber \\
	& &  - \frac{1}{3} Q \Big(E_{ii} -3 U_0\Big) \;\; \mathbb{I}_{F}[1;3/2] \Bigg] .
\end{eqnarray}
Here, we have $c_f=-1/2$ for fermions, and  $\tilde{b}_2^0=[U_f^2 -\frac{1}{3} {U_f}_{ii}  + \frac{1}{6} (G_{ij})^2]$, $\tilde{b}_{21}^T= \frac{1}{3} [ 2 E_i^2 + E_{ii0} - 3 {U_f}_{00}]$, $\tilde{b}_{22}^T= - \frac{2}{3} Q [E_{ii} -3 {U_f}_0]$. As $tr[U_f]=0$, we find $\tilde{b}_1$ will not contribute.  The other thermal HKCs are simplified further as $\tilde{b}_2^0=\frac{4}{3} \; [(G_{ij})^2+E_i^2-\frac{1}{3} {U_f}_{ii}]$. The effective action (\ref{EHL}) is the non-Abelian generalization of the Euler-Heisenberg effective action of QED \cite{Dunne:2004nc} at finite temperature.

Let us consider another scenario where heavy fermions are vector-like, and coupled with the light scalar, i.e., SM like Higgs \cite{DasBakshi:2020ejz, DasBakshi:2021ofj}. This UV theory possesses CP-violation, and once the heavy vector fermions are integrated out, CP-violating dimension-6  operators $(H^\dagger H \; G_{\mu \nu} \tilde{G}^{\mu \nu})$ generated. This CP violation may play a role in baryogenesis where it is known that the CP violation in the standard model CKM matrix is not enough and a new source of CP violation is required for generating successful baryogenesis \cite{Anderson:1991zb, Trodden:1998ym, Cline:2006ts}.


\section{ Coleman-Weinberg potential from Finite Temperature Effective Action}\label{CW-T2}
In this section, we derive the effective potential of the light scalars from the effective action at finite temperature (\ref{eq:eft_lag_finite_temp_boson}) and (\ref{eq:eft_lag_finite_temp_ferm}), which we obtained after integrating out heavy bosons or fermions respectively.

The effective action is the sum of the tree level action for the scalar for which we want the effective potential arising from the one loop effective actions at finite temperature,
\be
{\cal L}_{eff}(\phi_l,\beta)= {\cal L}_{tree}+{\cal L}_{l}^{(1)}(m_l,T)+{\cal L}_{h}^{(1)}(m,T).
\ee
 In the absence of gauge fields, the light field  effective action  after integrating out the heavy bosonic fields, in the absence of gauge interactions of the heavy bosons is from  (\ref{eq:eft_lag_finite_temp_boson}),
 \begin{eqnarray}\label{EFT-b1-h}
 {\cal L}_{h}^{(1)} = c_h\;\; \text{tr} \Big[ \mathbb{I}[0;0](m, T) &-& U \;\;\mathbb{I}[0;1] (m, T)+ \frac{1}{2}\big(U^2  \big)\;\; \mathbb{I}[0;2](m, T)\nn\\&-& \big(U^3\big) \;\; \mathbb{I}[0;3] (m, T)+\cdots \Big].
\end{eqnarray}    

Here, $U$ is defined from the tree-level UV interaction Lagrangian  as 
\be
U(\phi_l)= \frac{\delta^2 {\cal L}_{tree}(\phi_h,\phi_l)}{\delta \phi_h^2},
\ee
which is written in terms of $\phi_l$ using e.o.m of $\phi_h$. 
Similarly, we can write the one loop correction to the light field effective action from loops of light bosonic fields (of mass $m_l$) at finite temperature as
 \begin{eqnarray}\label{EFT-b1-l}
 {\cal L}_{l}^{(1)} = c_l\;\; \text{tr} \Big[ \mathbb{I}[0;0](m_l, T) &-& U_l \;\;\mathbb{I}[0;1](m_l, T) + \frac{1}{2}\big(U_l^2  \big)\;\; \mathbb{I}[0;2](m_l, T)\nn\\ &-& \big(U_l^3\big) \;\; \mathbb{I}[0;3](m_l, T)  +\cdots\Big],
\end{eqnarray}  
where $U_l$ is 
\be
U_l(\phi_l)= \frac{\delta^2 {\cal L}_{tree}(\phi_h,\phi_l)}{\delta \phi_l^2},
\ee
 written in terms of $\phi_l$ using e.o.m of $\phi_h$. In order to obtain a dimension-6 effective theory, we need to retain terms of order $U^3$ and $U_l^3$ in (\ref{EFT-b1-h}) and (\ref{EFT-b1-l}), respectively.

The Coleman-Weinberg \cite{Coleman:1973jx, Weinberg:1973am} effective potential, which captures the infrared dynamics of the light field, is obtained by integrating out the short-distance quantum fluctuations around a constant (space-time independent) background $\phi_b$.
To calculate the CW potential we  write the light scalar $\phi_l({\vec x}, \beta) =\phi_b + \varphi(\vec x, \beta)$ where $\phi_b$ is a constant background field,  and expand   the effective action in powers of quantum-fluctuations $\varphi({\vec x}, \beta)$, 
\be
{\cal L}_{eff}(\phi_l,\beta)=  {\cal L}_{eff} (\phi_b) + \frac{\delta {\cal L}_{eff}}{\delta \phi_l}\Big\vert_{\phi_l=\phi_b} \varphi({\vec x}, \beta)+\frac{1}{2} \frac{\delta^2 {\cal L}_{eff}}{\delta \phi_l^2}\Big\vert_{\phi_l=\phi_b} \varphi(\vec x, \beta)^2 + \cdots.
\label{CW-1}
\ee
The linear term is eliminated by using the e.o.m of $\phi_l$ in the classical effective action $ {\cal L}_{eff}(\phi_l,\beta)$. 
\be
e^{-V_{CW}(\phi_b, \beta)} \equiv \int D\varphi({\vec x}, \beta) e^{ - {\cal S}_{eff} (\phi_b) } \, e^{ -\int d^4 x  \frac{1}{2} (\delta^2 {\cal L}_{eff}/\delta \phi_l^2)\vert_{\phi_l=\phi_b} \varphi({\vec x}, \beta)^2}.
\ee
The second derivative of the action evaluated at constant background is the  effective mass of the light field, 
\be \label{meffCW}
m^2_{eff}(\phi_b) \equiv \frac{1}{2}\frac{\delta^2 }{\delta \phi_l^2}\left({\cal L}_{tree}+{\cal L}_{l}^{(1)}(m_l,T)+{\cal L}_{h}^{(1)}(m,T)\right)\Big\vert_{\phi_l=\phi_b} \,.
\ee
The effective mass-squared of light fields is the sum of three contributions:  tree level, one loop after integrating out light fields with mass $m_l$ and one loop contribution of integrating out the heavy fields with mass $m$.

For integrating out $\varphi({\vec x}, \beta)$, we can use the Heat-Kernel technique with $m_{eff}^2 (\phi_b)$ as the mass. Now  $U=0$ as the effective action is computed at a constant background. In the HK expansion, we have $b_0=1$ and $b_k=0$ for all $k=2,3,\dots$.
 We find that the Coleman-Weinberg potential from the effective action
 after integrating out the heavy fields at finite temperature is, using (\ref{eq:eft_lag_finite_temp_boson}) and  (\ref{I00-1}),  is given by,
\bea\label{CW-T}
V_{CW}(\phi_b,T)&=& c_s \;\;  \text{tr}[\tilde{b}_0]\;\; \mathbb{I}[0;0] (m_{eff}, T) \nn\\
&=&
\frac{c_s}{32\pi^2} \Bigg\{ m^4_{eff} \left(\ln\left(\frac{m^2_{eff}}{\mu^2}\right)-\frac{3}{2}\right)\Bigg\} - c_s\,\mathbb{S}[0;0]
(m_{eff}, T),
\eea
where the thermal correction to the one loop CW potential, which for $T> m/(2\pi)$ is given by the series  (\ref{S00}),
\bea
c_s\, \mathbb{S}[0;0] (m_{eff}, T)= \; \frac{\pi^2 T^4}{90} - \frac{ m_{eff}^2 T^2}{24} &+& \frac{m_{eff}^3 T}{12 \pi} \; +\frac{m_{eff}^4}{2\; (4\pi)^2}
  \Big\{ ln\Big[\frac{m_{eff} e^\gamma}{4\pi T} \Big] -\frac{3}{4}\Big \} \nn\\
  &-&\frac{\zeta(3) \; m_{eff}^6}{3 \; (4\pi)^4 T^2} +\cdots.
  \eea
The main ingredient in computing the thermal CW potential is calculation of the effective mass $m_{eff}$ at finite temperature in a given theory. For an effective theory, the general expression for the thermal corrected effective mass of the light field, using (\ref{EFT-b1-h}) and (\ref{EFT-b1-l}), can be written explicitly as
\bea\label{meffT}
{m_{eff}^2}^{(1)}(\phi_b,T)&=&m_l^2 + \frac{1}{2} U_l(\phi_l) \nn\\
&+& \frac{1}{2}\,\,c_l\; \text{tr} \Big[ -\big( U_l(\phi_l))^{\prime\prime} \;\;\mathbb{I}[0;1] (m_l,T)+ \frac{1}{2}\big(U_l^2 (\phi_l) \big)^{\prime \prime} \;\; \mathbb{I}[0;2](m_l,T) \nn\\
&-& \big(U_l^3(\phi_l)\big)^{\prime \prime} \;\; \mathbb{I}[0;3](m_l,T) \Big]\Bigg\vert_{\phi_l=\phi_b} \nn\\
&+&\frac{1}{2}\,\,c_h\; \text{tr} \Big[ -\big( U(\phi_l))^{\prime\prime} \;\;\mathbb{I}[0;1] (m,T)+ \frac{1}{2}\big(U^2 (\phi_l) \big)^{\prime \prime} \;\; \mathbb{I}[0;2](m,T) \nn\\
&-& \big(U^3(\phi_l)\big)^{\prime \prime} \;\; \mathbb{I}[0;3](m,T) \Big]\Bigg\vert_{\phi_l=\phi_b},
\label{meff}
\end{eqnarray}    
where primes acting on $U$ denotes derivative {\it w.r.t} the light field $\phi_l$. This is the effective mass, including the one loop thermal corrections from the heavy fields and light fields, and this is substituted in (\ref{CW-T}) to obtain the finite temperature Coleman-Weinberg potential for an effective theory.

We summarise some of the salient features of the construction of the Coleman-Weinberg potential of a scalar at finite temperature in an effective theory obtained by integrating out the heavy fields:

\begin{itemize}

\item
The thermal contribution to the effective action from the integrated out heavy fields which are represented by terms
of dimension-4 and dimension-6, $ \mathbb{I}[0;2](m, T)$, and $ \mathbb{I}[0;3](m, T)$ respectively have not been considered in the earlier literature \footnote{These terms, without encapsulating the contribution from Polyakov loop, up to dimension six is computed employing diagrammatic method in Ref.~\cite{Chala:2024xll}} and are essential for a consistent calculation of the thermal effective action and the subsequent calculation of the thermal CW potential from such theories.

\item
The thermal corrections  to the light field effective mass from  dimension-4 and dimension-6  operators $(U_l^2)^{\prime \prime}$ and $(U_l^3)^{\prime \prime}$ which are given by  
$ \mathbb{I}[0;2](m_l, T)$ and $ \mathbb{I}[0;3](m_l, T)$ have not been considered in the literature earlier when computing the thermal CW potential of effective theories.

\item
When we gauge interactions of the heavy fields, we find contributions in two ways. The HK coefficients $\tilde b_i, i=0,1,2,\cdots $ generalise to include terms with gauge fields $G_{0i}, G_{ij}$ and gradients of spatial and temporal gradients of the potential $U_i, U_0$, etc. 
There is also a contribution from the Polyakov loops $\Omega_l$ and $\Omega_h$.
Here $\Omega_l$ depends on the gauge charges of the light fields. In the case of Higgs potential, $\Omega_l$ will arise from the $SU(2)$ charge of the Higgs. The Polyakov loop contribution of the heavy fields that have been integrated out will be present in $\Omega_h$, which will depend on the gauge representations of the respective heavy fields.

\item
Once we have obtained a light field EFT, we construct the effective Coleman-Weinberg potential by integrating out the short wavelength fluctuations of the light fields. In deriving (\ref{CW-T}), it is required that $m_{eff} \simeq U_l(\phi_b)  $ is the larger than all the light field masses $m_l$  in the theory effective theory, as $m_{eff}$ acts as the infrared infrared regulator \cite{Barvinsky:2002uf}. 

For this procedure to be valid, the tree level light field masses $m_l$ must obey the constrain,
\be
m_l^2 \leq U_l(\phi_b).
\label{LFC}
\ee
The condition (\ref{LFC})  gives us a criterion of which fields to treat as light and which are to be treated as heavy fields with mass $m$ discussed in Section~\ref{EFT}. For example, for evaluating the CW potential of the SM Higgs $h$ we count the particles with masses $m_i$ ( like top and $W, Z$) as light as their masses obey the condition
$m_i^2 < -\mu^2 + \frac{\lambda}{2} h^2  $
while particles with masses $m$  which are heavier than the field dependent effective Higgs mass
$m^2 > -\mu^2 + \frac{\lambda}{2} h^2  $
must be first integrated out using the method of Section~\ref{EFT} to derive the effective action at a finite temperature of the light Higgs, and then the CW potential can be calculated of the effective Higgs theory.

\item 
In deriving the effective theory at finite temperature (\ref{eq:eft_lag_finite_temp_boson}) and (\ref{eq:eft_lag_finite_temp_ferm}), the only assumption made is $ m > m_l$. The temperature can follow any of the three hierarchies  $T>m > m_l$, $T<m_l < m$ or $m> T > m_l$. The expressions for the $ \mathbb{I}[k;l] (m, T)$, which we reduce to sums of modified Bessel functions $K_\nu(n m \beta)$, remain the same irrespective of the temperature hierarchy. One can then use different asymptotic expansions of $K_\nu(n m \beta)$ for $ m \beta > \pi/2$ or $ m \beta < \pi/2$ to obtain the high or low-temperature expansions, respectively, depending on the application at hand.

\end{itemize}

In the next sections, we give examples of singlet heavy scalars that are integrated out to obtain the Higgs-EFT at finite temperature and the Coleman-Weinberg potential from this EFT. We elaborate on the effect of Polyakov loop contributions in the Coleman-Weinberg potential, which, to our knowledge, has not been studied earlier, and examine their role in determining the nature of phase transitions.


\section{Phase transitions in effective theories} \label{PhaseT}
In this section, we will illustrate the calculation of phase transitions in effective theories, which follows from the effective potential at finite temperature worked out in Section~\ref{CW-T2}. In this section, we will set the contributions of Polyakov loops and background gauge fields to zero. We study in detail the effect of Polyakov loops in phase transition in Section~\ref{PL}.

Specifically working in SMEFT, it was shown in \cite{Postma:2020toi} that phase transitions calculations from effective theories capture only a small slice of the parameter space requiring a first-order phase transition (FOPT) compared to the calculation from the full UV complete theory \cite{Grojean:2004xa, Bodeker:2004ws, Delaunay:2007wb, Chung:2012vg, Blinov:2015vma, Cline:2021iff, Damgaard:2015con,deVries:2017ncy}. To see why this is so,  we start with the SMEFT for the Higgs boson $H=\frac{1}{\sqrt{2}}(0, h)^T$ up to dimension-6 given by
\be\label{dim6}
{\cal L}_{eff}^{(6)}= \frac{1}{2} (\partial h)^2 - \left( \frac{1}{2} a_2 h^2 + \frac{1}{4} a_4 h^4 +\frac{1}{6} a_6 h^6 \right).
\ee
We fix some of the parameters  on the zero temperature potential $V_0(h)$ from (\ref{dim6}) by imposing the conditions
\be
\partial_h V_0(h)\Big\vert_{h=v}=0, \quad \quad \partial^2_h V_0(h)\Big\vert_{h=v}= m_{h0}^2 =(125 {\rm Gev})^2,
\ee
with $v=246 {\rm GeV}$. Imposing these conditions gives us $a_2=\frac{1}{2}\left(m_{h0}^2 -2 a_6 v^4 \right)$ and $a_4=\left(\frac{m_{h0}^2}{2v^2} -2 a_6 v^2 \right)$ and we can express the potential as
\be
V_0(h)= - \frac{1}{4}\left(m_{h0}^2 -2 a_6 v^4 \right)h^2+ \frac{1}{4} \left(\frac{m_{h0}^2}{2v^2} -2 a_6 v^2 \right)h^4 +\frac{1}{6} a_6 h^6,
\ee
The tree level effective Higgs mass is
\bea
m_h^2 &=&  \partial^2_h V_0(h) = a_2 + 3 a_4 h^2 +5 a_6 h^4 \nn\\
&=& -\frac{1}{2} m_{h0}^2 + a_6 v^4 + \frac{3}{2} \left(\frac{m_{h0}^2}{v^2} -4 a_6 v^2 \right)h^2 +5 a_6 h^4.
\eea
When the heavy sector masses are very heavy compared to the SM Higgs mass and the temperature hierarchy is $ m \gg T >m_l$, then the thermal corrections to Higgs mass from integrating out the heavy sector (\ref{EFT-b1-h}) are exponentially suppressed by the factor $e^{-m/T}$ and we can drop these terms. The thermal corrections to the Higgs mass will come from the SM fields in (\ref{EFT-b1-l}). Following this assumption, we can write the thermal effective potential as a power of $h$ as
\bea 
V(h,T)&=&V_0(h) + V_l(m_l, T)\nn\\
&\equiv& \frac{1}{2}\tilde a_2(T) h^2 + \frac{1}{2 \sqrt{2}} \tilde a_3(T) h^3  + \frac{1}{4} \tilde a_4(T) h^4 + \cdots 
\label{VhT}
\eea
In the SMEFT, the total tree level and loop contributions to the coefficients are
\bea\label{aiT}
\tilde a_2(T)&=& -\frac{1}{2}\left(m_{h0}^2 -2 a_6 v^4 \right) + T^2 \left(\frac{1}{4} \Big(\frac{m_{h0}^2}{2v^2} -2 a_6 v^2 \Big) + \frac{3}{16} (3 g^2 + {g^\prime}^2 ) + \frac{1}{4} y^2 \right)\,,\nn\\
 \frac{1}{2 \sqrt{2}}\tilde a_3(T)&\equiv&- \frac{1}{2 \sqrt{2}} E T = -\frac{1}{48 \pi}T \left( 3 g^3 + \frac{3}{2}(g^2 +{g^\prime}^2 )^{3/2} +  12\sqrt{3}\Big(\frac{m_{h0}^2}{2v^2} -2 a_6 v^2 \Big)^{3/2} \right) \,,\nn\\
\tilde a_4(T)&\simeq& \Big(\frac{m_{h0}^2}{2v^2} -2 a_6 v^2 \Big).
\eea
The $\tilde a_3$ coefficient, which is negative and which is crucial for phase transitions, arises from gauge boson and Higgs loops and is dominated by the gauge boson contribution
\be
E \simeq \frac{1}{12 \sqrt{ 2} \pi}  \left( 3 g^3 + \frac{3}{2}(g^2 +{g^\prime}^2 )^{3/2} \right).
\ee
At the transition temperature $T_c$ we must have $V(h=v_c, T=T_c)=0$. This phenomenon can happen when the cubic term and quartic terms in (\ref{VhT}) cancel among themselves as the quadratic term is smaller than each of these terms for $h\gg v$. Solving $V(h=v_c, T=T_c)=0$ for $T_c$ we have
\be
R_c \equiv \frac{v_c}{T_c} = \frac{\sqrt{2}  E}{\tilde a_4}= \frac{1}{12 \pi}\frac{ \left( 3 g^3 + \frac{3}{2}(g^2 +{g^\prime}^2 )^{3/2} \right)}{\Big(\frac{m_{h0}^2}{2v^2} -2 a_6 v^2 \Big)}.
\label{Rc}
\ee
The condition for the first-order phase transition is $R_c >1$. If we have only the standard model without the dimension 6-terms, then setting $a_6=0$ in (\ref{Rc}), we have
\be \label{RcSM}
R_c =\frac{v^2}{6 \pi m_{h0}^2} \left( 3 g^3 + \frac{3}{2}(g^2 +{g^\prime}^2 )^{3/2} \right).\ee
Taking the weak coupling $g^2/(4\pi)=1/29.5$, we see that the condition $R_c>1$ in (\ref{RcSM}) can be met only if the Higgs mass $m_{h0} < 31.5$ GeV. Therefore, in the standard model, we do have an FOPT.  

In SMEFT, keeping the contribution of the dimension-6 term in (\ref{Rc}), we see that the condition for FOPT,  $R_c>1$, can be written the constraint on $a_6$ given by
\be\label{a6}
\Big(\frac{m_{h0}^2}{2v^2} -2 a_6 v^2 \Big) < \frac{1}{12 \pi}\left( 3 g^3 + \frac{3}{2}(g^2 +{g^\prime}^2 )^{3/2} \right) \simeq 0.256.
\ee
The dimension-6 contribution requires to be fine-tuned so that the denominator in (\ref{Rc}) is small and the parameter space (\ref{a6}) for FOPT is therefore limited. 

    The contribution of the dimension-6 operators is significant when the heavy particle mass scale 
    $a_6 \sim m^{-2} \simeq (800\, {\rm GeV} )^{-2} $ \cite{Grojean:2004xa, Bodeker:2004ws}. This implies that for temperatures $ T>m/(2\pi)$, the heavy particle thermal loop contributions from (\ref{EFT-b1-h}) must be included in the light particle thermal potential (\ref{aiT}).
 Heavy bosons will contribute to the cubic term in (\ref{aiT}), and this will give us extra parameter space for achieving FOPT. We illustrate this with an example of Higgs coupling with a heavy scalar and the contribution of the integrated out scalar in the Higgs phase transition.


\subsection{Heavy scalar UV theory}
We consider a UV theory with renormalizable coupling terms with a light scalar $\phi$ and a heavy scalar $\Phi$ of mass $m$ given by
\be
V_{UV}= m^2 \Phi^2+\frac{1}{2}\lambda_h \Phi^4 + \mu_1 \phi^2 \Phi +\lambda_1 \phi^3 \Phi +\lambda_{hl}\phi^2 \Phi^2 + \mu_2 \phi \Phi^2 +\frac{1}{2} \lambda_l \phi^4 + \mu_l^2 \phi^2\,\,.
\ee
The lagrangian of the heavy field can be written as
\be
{\cal L}_{UV}= \Phi \left( P^2 -m^2 -U \right)\Phi + B \Phi +h.c.
\label{UV}
\ee
where 
\be
U =   \lambda_h \Phi^2 + \lambda_{hl}\phi^2 +\mu_2 \phi,
\label{U}
\ee
and 
\be
B= \mu_1 \phi^2 + \lambda_1 \phi^3\,\,.
\label{B}
\ee
We need to integrate out the heavy field $\Phi$ and obtain the effective potential in terms of the light field $\phi$. We can use the e.o.m of $\Phi$ to write $U$ in terms of the light fields as follows. e.o.m of $\Phi$ is given by
\be
 \left(P^2 -m^2 -U\right)\Phi=-B,
\ee
which can be expanded treating $m^2$ as the heavy scale as
\bea \label{eom}
\Phi&=&-\frac{1}{ \left(P^2 -m^2 -U\right)}B=\frac{1}{m^2}\left(1+\frac{(P^2-U)}{m^2} + \frac{(P^2-U)^2}{m^4} +\cdots \right) B \nn\\
&\simeq& \frac{1}{m^2} B.
\eea
Using this, we can express (\ref{U}) in terms of the light fields as
\bea
U =
 \frac{\lambda_h}{m^4} \left(\mu_1 \phi^2 + \lambda_1 \phi^3 \right)^2+ \lambda_{hl}\phi^2 +\mu_2 \phi
\label{U2}
\eea
The light field effective mass  at tree level $U_l$ is 
\be
U_l(\phi)= \mu_l^2 \phi^2+ \lambda_l \phi^4  + \frac{\mu_1^2}{ m^2}\left( \mu_1 \phi^2 +  \lambda_1 \phi^3\right) + \frac{\lambda_{hl}} {m^4} \left( \mu_1 \phi^2 +  \lambda_1 \phi^3\right)^2.
\ee
The heavy field effective mass at the tree level, which is used in (\ref{EFT-b1-h}), is
\be
m^2_{eff} \equiv \frac{1}{2}\frac{\delta^2 V_{UV}}{\delta \Phi^2}\Big\vert_{\Phi=B/m^2}.
\ee

One can compute the effective action more precisely by including the effects of loops consisting of light-heavy scalar propagators. In that case, we need to work with complete action as we can not disentangle the free part from the interaction part, see Ref.~\cite{Banerjee:2023xak}. 
Here, we are providing the sample structures of a few relevant terms :
\begin{eqnarray}
	\mathcal{L}_{eff} & \supset &  \frac{C_s}{(4\pi)^2} \Big[C_{i} tr[U_{i}] + C_{ij} tr [U_{ij} U_{ji}] + C_{ijk} tr [U_{ij} U_{jk} U_{ki}] \Big],
\end{eqnarray} 
where $i,j,k \in \{1,2,..,n \}$ for $n$-number of non-degenerate scalar fields. Here, $U_{ij}=\frac{\delta^2\mathcal{L}}{\delta \Phi_i \delta \Phi_j}$ with $\Phi_i$ is the scalar field with mass $M_i$. For $n=2$, a few sample $C_i$'s can be given as 
\cite{Banerjee:2023xak},
\begin{eqnarray}
	C_1 & = &  M_1^2 + ln (M_1^2/\mu^2);  C_{12} = 1- \frac{M_1^2}{\Delta_{12}^2}  ln (M_1^2/\mu^2) +  \frac{M_2^2}{\Delta_{12}^2}  ln (M_2^2/\mu^2);   \nonumber \\ 
	C_{122} & =& \frac{1}{\Delta_{12}^2} - \frac{M_1^2}{(\Delta_{12}^2)^2}   ln (M_1^2/M_2^2).
\end{eqnarray}


\subsection{SM Higgs + heavy singlet }
To illustrate the general formalism developed in the earlier sections, we consider the UV theory of SM Higgs interacting with a heavy singlet scalar $\Phi$ \cite{Anderson:1991zb}. The general gauge invariant potential for the Higgs scalar $\langle H^T\rangle =(1/\sqrt{2})(0,h)$  with scalar singlet is
\be \label{singlet}
V_{UV}= \frac{m^2}{2}\Phi^2+\lambda_h \Phi^4 + \frac{1}{2}\mu_1 h^2 \Phi + \frac{1}{2}\lambda_{hl}h^2 \Phi^2+ \frac{1}{4} \lambda_l h^4 - \frac{1}{2}\mu_l^2 h^2\,\,, \quad \quad \mu_l^2 >0\,.
\ee
To keep the discussion simple, we will impose an additional $Z_2$ symmetry on $\Phi$, which makes $\mu_1=0$. We will take $m^2 > 0$ and $\lambda_h>0$, and the heavy scalar does not take a vacuum expectation value at zero temperature. The field-dependent mass terms of the heavy and light fields  are, therefore
\bea
U(h) =m^2 +\lambda_{hl}h^2,
\label{U2}
\eea
and
\be
U_l(h)= -\mu_l^2 + \frac{1}{2} \lambda_l h^2.
\ee

To be sensitive to up to dimension-6, we need to keep terms up to $U^3$ and $U_l^3$ order 
in (\ref{EFT-b1-l}), which would be used for computing the effective thermal masses (\ref{meffT}), which in turn will be used for computing the Coleman-Weinberg potential  (\ref{CW-T}). 

Starting from a UV theory, we can integrate out the heavy field $\Phi$ to obtain the effective one loop potential at zero temperature at a lower scale $\mu$ given by
\bea
V_{one loop}(h)= &&\frac{c_s}{32 \pi^2} \Bigg\{ m^4 \left(\ln\left(\frac{m^2}{\mu^2}\right)-\frac{3}{2}\right)+ m^2 \left(1-\ln\left(\frac{m^2}{\mu^2}\right)\right) U\nn\\
 &-&\frac{1}{2} \ln\left(\frac{m^2}{\mu^2}\right) U^2 +\sum_{n=3}^{\infty} \frac{(-1)^n}{n!} m^{4-2n} U^n\Bigg\}.
\label{1loop}
\eea

The quadratic and quartic couplings of the Higgs field $h$ in   the effective potential 
\be
V(h)= V_{tree}(h) +V_{one-loop}(h),
\ee
where 
\be \label{BC1}
V_{tree}(h)= \frac{1}{4} \lambda_l h^4 - \frac{1}{2}\mu_l^2 h^2\,,
\ee
 are fixed using the following conditions,
\be
\frac{\partial V(h)}{\partial h}\Big\vert_{h=v, \mu=v}=0\,,\quad  \frac{\partial^2 V(h)}{\partial h^2}\Big\vert_{h=v, \mu=v}=m_{h0}^2,
\ee
with $v=246 {\rm GeV}$ and $m_{h0}=125.7 {\rm GeV}$.
Imposing the boundary conditions (\ref{BC1}) at zero temperature, we can write the quadratic and quartic couplings in terms of the observable parameters $m_{h0}$ and $v$. The $\log\mu$ terms are absorbed in these observable parameters \cite{Kanemura:2022txx, Hashino:2022ghd}. Keeping terms up to dimension-6 in the Higgs field, we can write the Higgs potential at zero temperature as
\be
V(h)= - \frac{1}{4}\left(m_{h0}^2 -2 a_6 v^4 \right)h^2+ \frac{1}{4} \left(\frac{m_{h0}^2}{2v^2} -2 a_6 v^2 \right)h^4 +\frac{1}{6} a_6 h^6,
\ee
with $a_6= -\frac{1}{6 M^4} $.

In order to calculate the finite temperature potential effective potential for the Higgs, we need to include contributions from the light fields $(h, W, Z, t)$ as well as the thermal contribution from $\Phi$ loops given in  (\ref{EFT-b1-h}) with the effective mass of $\Phi$ taken as $m_{\Phi}^2=m^2 + \lambda_{hl} h^2$. We will then get additional contributions to the thermal Higgs potential (\ref{VhT}),
\bea 
V(h,T)&=&V_0(h) + V_l(h, T)+  + V_h(h, T)\nn\\
&\equiv& \frac{1}{2}\tilde a_2(T) h^2 + \frac{1}{2 \sqrt{2}} \tilde a_3(T) h^3  + \frac{1}{4} \tilde a_4(T) h^4 + \cdots 
\label{VhT2}
\eea
Now, the total tree level and loop contributions from the heavy and light fields to the coefficients are
\bea\label{aiT2}
\tilde a_2(T)&=& -\frac{1}{2}\left(m_{h0}^2 -2 a_6 v^4 \right) + T^2 \left( \frac{1}{4}\Big(\frac{m_{h0}^2}{2v^2} -2 a_6 v^2 \Big) + \frac{3}{16} (3 g^2 + {g^\prime}^2 ) + \frac{1}{4} y^2  +\frac{1}{24}\lambda_{hl} \right)\,,\nn\\
\frac{1}{2 \sqrt{2}}\tilde a_3(T)&\equiv&- \frac{1}{2 \sqrt{2}} E T = -\frac{1}{48 \pi}T \left( 3 g^3 + \frac{3}{2}(g^2 +{g^\prime}^2 )^{3/2} +12\sqrt{3}  \Big(\frac{m_{h0}^2}{2v^2} -2 a_6 v^2 \Big)^{3/2} +4\lambda_{hl}^{3/2}\right) \,,\nn\\
\tilde a_4(T)&\simeq& \Big(\frac{m_{h0}^2}{2v^2} -2 a_6 v^2 \Big).
\eea
where we assumed $m^2 +\lambda_{hl}h^2 \simeq \lambda_{hl}h^2$ which can be valid for $m$ not to large compared  to $v$ and $\lambda_{hl}\sim1$ in the range of the field  $h\sim v_c \gg v$.

The critical temperature for the phase transition is the temperature at which the potential has two degenerate minima separated by a potential hump,
\bea \label{VTc}
\frac{\partial V(h, T=T_c)}{\partial h}\Big\vert_{h=v_c}=0\,,\quad V(h=0,T_c)=0\,,\quad V(v_c,T_c)=0\,.
\eea

Computing the  critical temperature using (\ref{VTc}) we find
\be
R_c \equiv \frac{v_c}{T_c}  \frac{\sqrt{2} E}{ \tilde a_4}\simeq \frac{1}{12 \pi}\frac{ \left( 3 g^3 + \frac{3}{2}(g^2 +{g^\prime}^2 )^{3/2} +4\lambda_{hl}^{3/2}\right)}{\Big(\frac{m_{h0}^2}{2v^2} -2 a_6 v^2 \Big)}.
\label{Rc2}
\ee
We see that due to heavy-light mixing coupling $\lambda_{hl}$, we have more parameter space for getting $R_c >1$ and a first-order phase transition. If we had started with the dimension-6 effective theory (\ref{dim6}) at zero temperature and included thermal loop corrections from the SMEFT particles, we would obtain $R_c$ given in (\ref{Rc}) and missed the thermal contribution coming from heavy-light mixing term. 

In conclusion, this section has illustrated that the correct procedure to compute effective potential at finite temperature in EFTs is to derive the thermal EFT from a UV theory instead of starting with a zero-temperature EFT and computing thermal corrections from the light particles only.


\section{Polyakov loop contribution to effective potential and phase transitions}\label{PL}
In this section, we will focus on the contribution of Polyakov loops in the Coleman-Weinberg effective potential at finite temperature and study the effect of PLs on phase transitions. For the computation of the Higgs effective potential, Polyakov loops can arise in two ways:
\begin{itemize}
\item PL contribution to Higgs potential can arise from the weak $SU(2)$ gauge charge of the Higgs, or in the Higgs carries non-zero charges of a dark sector gauged $SU(N)$.

\item PL terms in the Higgs potential can also arise from heavy scalar or fermions that have been integrated out to give the SMEFT if those heavy particles carry non-zero charges of some gauged $SU(N)$ group. These contributions are important at high temperatures $T>m$ when these heavy particles will be present in the thermal plasma.
\end{itemize}

The finite temperature effective potential arising from a  scalar loop is given by (\ref{S00PL}),
\bea
{V}_{b}^{one-loop}&=&   -\frac{1}{2}\mathbb{I}[0,0]=\frac{m^4}{64 \pi^2} \Big(\ln\Big(\frac{4\pi m^2}{\mu^2}\Big)+\frac{3}{2}-\gamma_E\Big)-\frac{m^4}{32 \pi^2 }\left(\log \Big(\frac{\beta  m e^{\gamma_E}}{4 \pi}\Big)
-\frac{3}{4}+  {\tilde n}^2 \zeta (3) \right)\nn\\
\nn\\
&-& \frac{\pi ^2 \left(1-30 {\tilde n}^2\right)}{90 \beta ^4} 
+
\frac{\left(1+6 {\tilde n}^2\right) m^2}{24 \beta ^2}-\frac{2 \pi  {\tilde n}^2 m}{15 \beta ^3}-\frac{m^3}{12 \pi  \beta }+\frac{ \beta ^2 m^6 \left(6 {\tilde n}^2 \zeta (5)+\zeta (3)\right)}{384 \pi ^4} +\cdots \nn\\
\eea
 we see that Polyakov loops make a negative contribution to the pressure $p=\frac{\pi ^2 \left(1-30 {\tilde n}^2\right)}{90 \beta ^4} +{\cal O}(m \beta).$
 
 Moreover, that may possess significance for cosmological applications. If there are gauged scalar particles in the thermal bath with masses $m<T$, they will contribute to the cosmological Hubble expansion factor as negative effective neutrino degrees of freedom $ \Delta N_{\nu}=  - \frac{4}{7} \, 30 {\tilde n}^2$.

 The negative $ m^3 T$ term, which is important for phase transitions, is unaffected by the contribution of the Polyakov loop. However, novel terms like $m T$ are absent without Polyakov loops, and extra contributions to $ m^2, m^4$, and $ m^6$ coefficients from Polyakov loops will change the parameter space for phase transitions. 

Polyakov loop contributions also arise from fermions in the heat bath; for example, the Higgs potential will receive  Polyakov loop corrections from top quark loops, from the strong interaction $SU(3)$ and from other possible dark-sector gauged interactions of the top quark. The effective scalar potential arising from gauged fermions in the loop is from (\ref{PLFL-1}) given by,
 \bea\label{PLFL-2}
{V}_{f}^{one-loop}&=&-c_f\, \text{tr}\; [ \tilde{b}_0\;\mathbb{I}_{\Omega F}(0,0)] \\
&=& (-4) \Bigg[ \frac{m^4}{64 \pi^2} \Big(\ln\Big(\frac{4\pi m^2}{\mu^2}\Big)+\frac{3}{2}-\gamma_E\Big)  -\frac{m^4 }{64 \pi ^2} \left(ln \Big(\frac{\beta m}{4\pi}\Big)-\frac{3}{4}+ \tilde n^2 \zeta(3) \right) \nn \\
&-&\frac{7}{8}\frac{ \pi ^2 \left(30 {\tilde n}^2-1\right)}{90 \beta ^4}
-\frac{\left(6 {\tilde n}^2+1\right) m^2}{48 \beta ^2}+\frac{\pi  {\tilde n}^2 m}{10 \beta ^3} +\frac{7 \beta ^2 m^6 \left(6 {\tilde n}^2 \zeta (5)+\zeta (3)\right)}{3 (2 \pi )^4}+\cdots \Bigg]. \nn
\eea
We would like to emphasize that for Chiral fermion in (3+1) dimension,  $\text{tr}(\tilde{b}_0)=4$. There are contributions to the $m^2, m^4$ and $m^6$  terms in the potential and a novel linear $mT^3$ term. Polyakov loops make a negative to the pressure, which is equivalent to  $\Delta N_\nu= - 15 {\tilde n}^2 $ extra neutrino species.

The expressions (\ref{S00PL}) and (\ref{PLFL-2}) are  perturbative expansions in $\tilde n$ valid for $\tilde n < 1/(2\pi)$.  For the value of $\tilde n \sim 1$, one must do a non-perturbative summation of the Bessel series of $\mathbb{I}(0,0)$ in (\ref{non-pert-PL}). In numerical computations of phase transitions, one can use the expression  (\ref{non-pert-PL}) valid for all values of $m \beta$ and the entire range of ${\tilde n} \in [-1,1]$ for an accurate calculation of the parameters.

\section{Conclusions}\label{Conclusions} 
Effective actions at finite temperatures have many applications in colliders, in the study of quark-gluon plasmas, and in the study of non-perturbative phenomena like the Schwinger mechanism and Casimir effect. Effective theories also have cosmological applications, like calculations of the relic density of dark matter and phase transitions. 

The standard approach starts with an SMEFT of some suitable dimension $n>4$ and uses the standard calculations of finite temperature field theory. This approach breaks down for $T > m$, i.e., when the application involves temperatures larger than the mass of the heavy fields integrated out. This is so because when $T>m$, the thermal loops have the heavy particles, but the finite temperature calculations starting from an SMEFT can only have the SM particles in the loops. This lacuna of the standard method of using SMEFT at finite temperatures is seen in the computation of phase transitions in effective theories, where explicit calculations show that the parameter space of phase transitions computed with the full UV theories do not match the results of calculations using SMEFT.

In this paper, we use the Heat-Kernel method to compute the finite temperature Wilson coefficients of the higher dimension operators. The zero-temperature HK method gives a systematic expansion of higher dimension operators as an expansion in the inverse powers of the heavy field mass, which is integrated out. This is particularly useful when the 
heavy fields have gauged charges, and the HK method provides the higher dimension operators, which are covariant functions of the corresponding 'electric' and 'magnetic' fields and their derivatives. In the finite temperature calculation of the HK coefficients, we obtain the thermal factors ${\mathbb I}(k,l) $ associated with the operators at all orders. We compare our results with the known results; for example, the ${\mathbb I}(0,0)$ gives the pressure, and ${\mathbb I}(0,1)$ gives the thermal mass of bosons and fermions. We give the general expressions for calculations of ${\mathbb I}(k,l) $ for all operators and as perturbative expansions in $m \beta$. These results are then used in the calculations of the finite temperature Coleman-Weinberg potential of higher dimension effective theories, which are used in the phase transition calculations. Of particular interest is our calculation of the contribution of Polyakov loops in the effective action at finite temperature and the Coleman-Weinberg potential. We also show that the PL contributions can change the parameter space for phase transitions.

In this paper, we develop the formalism and give the general results. In future work, we will apply the results to compute phase transitions in specific models.

\acknowledgments{ JC thanks Sabyasachi Chakraborty, Diptarka Das, Apratim Kaviraj, Kaanapuli Ramkumar, and Nilay Kundu for useful discussions. JC acknowledges support from CRG project, SERB, India. SM thanks the IIT-Kanpur  grant of a Distinguished Visiting Professor position,  for support in carrying out this work}.

\appendix

\section{Appendix}\label{Appendix}

\subsection{Thermal Heat-Kernel coefficients}\label{HKC}
Here, we provide some of the relations explicitly:
\begin{eqnarray}
\text{I.}\;\;\;\; 	k_1=0 & \implies & k_3=k_2=0:   b_0 = \tilde{b}_0 = 1.\\ \nonumber \\
\text{II.}\;\;\;\;    	k_1=1 & \implies & k_3=1,k_2=0; \nonumber \\ \nonumber \\
	& &  k_3=0,k_2=1: \nonumber \\ 
	-b_1 = Q^2\tilde{b}_0 - \tilde{b}_1 &\implies & \tilde{b}_1 = Q^2 b_0 + b_1 = Q^2 + U - Q^2= U.\\ \nonumber \\
\text{III.}\;\;\;\;	k_1=2 & \implies & k_3=2,k_2=0; \;\; k_3=1,k_2=1;\;\; k_3 = 0, k_2 = 2.\nonumber \\
	b_2/2! & = & \frac{(Q^2)^2}{2!}\tilde{b}_0 - Q^2 \tilde{b}_1 + \frac{\tilde{b}_2}{2!} \nonumber \\
	\tilde{b}_2 & =&  b_2 - (Q^2)^2 \tilde{b}_0 + 2 Q^2 \tilde{b}_1 \nonumber \\
	& = & (U-Q^2)^2 - (Q^2)^2  + 2 Q^2 U - \frac{1}{3} (U-Q^2)_{ii} + \frac{1}{6} (G_{ij})^2 \nonumber \\
	& = & U^2 -\frac{1}{3} U_{ii} + \frac{1}{3} (Q^2)_{ii} + \frac{1}{6} (G_{ij})^2+ [Q^2, U].  \\ \nonumber \\
\text{IV.}\;\;\;\;	k_1=3 & \implies & k_3=3,k_2=0; \;\; k_3=2,k_2=1; \;\; k_3 = 1, k_2 = 2; \;\; k_3 = 0, k_2 = 3.\nonumber \\
	- b_3/3! & = & \frac{(Q^2)^3}{3!}\tilde{b}_0 - \frac{(Q^2)^2}{2!\; 1!} \tilde{b}_1 + \frac{(Q^2)}{1! \;2!} \tilde{b}_2 - \frac{\tilde{b}_3}{3!} \nonumber \\
	\tilde{b}_3 &=&    b_3  + (Q^2)^3 \;\tilde{b}_0 - 3 (Q^2)^2 \tilde{b}_1 + 3 (Q^2) \tilde{b}_2 \nonumber \\
	& =& U^3 - 2 [Q^2,U] U + U [ Q^2,U] + [Q^2, [Q^2,U]] - Q^2 U_{ii} + Q^2  (Q^2)_{ii} \nonumber \\ 
	& + &  \frac{1}{2} Q^2 (G_{ij})^2 + \frac{1}{2} (U_i -Q^2_i)^2 + \frac{1}{2} (U-Q^2) (G_{ij})^2  + \frac{1}{10} (J_i)^2 - \frac{1}{15} G_{ij} G_{jk} G_{ki} \nonumber \\
	& + & \frac{1}{10} (U-Q^2)_{iijj} -\frac{1}{30} [D_i, [D_j,[D_j,J_i]]].
\end{eqnarray}
Now, at this point, we have to be careful about the distinction between $Q$ and $D_0$. $Q$ is a c-number only when brought to the extreme left of any given term while within the commutator the operator $D_0$ will play a central role as $Q= I p_0 +D_0$. Keeping this in mind we will compute the the following commutator as follows

\begin{eqnarray}
	[Q^2, U] & = & Q[Q,U] + [Q,U]Q  =  Q U_0 + U_0 Q \nonumber \\
	& = & Q U_0 + QU_0 - [Q,U_0]  =  2QU_0 -U_{00},
\end{eqnarray}
where $U_0 = [D_0,U] = (D_0 U)\equiv [Q,U], U_{00}= [D_0^2, U]= [D_0, [D_0, U]] \equiv [Q,U_0]$.
\begin{eqnarray}
	UQ &=& -[Q, U] + QU	 =  QU -U_{0}.
\end{eqnarray}
\begin{eqnarray}
	[Q, [Q^2, U]] & = & [Q,2QU_0 -U_{00}]   =  2QU_{00} -U_{000}.
\end{eqnarray}
\begin{eqnarray}
	[Q^2, [Q^2, U]] & = & [Q^2,2QU_0 -U_{00}]   =  4Q^2 U_{00} - 4Q U_{000}  -  U_{0000}.
\end{eqnarray}
\begin{eqnarray}
	U_{000}Q &=& -[Q, U_{000}] + QU_{000}  =  QU_{000} -U_{0000}.
\end{eqnarray}
Note that 
\begin{eqnarray}
	[D_0, [D_0, U]] & = & [D_0, (D_0 U - U D_0)] = D_0^2 U -D_0 U D_0 + D_0 U D_0 - U D_0^2; \nonumber 
\end{eqnarray}
\begin{eqnarray}
	[ D_{0}^2, U ] & = & D_0 [D_0 U - U D_0] + [D_0 U - U D_0] D_0  = D_0^2 U -D_0 U D_0 + D_0 U D_0 - U D_0^2 \nonumber \\ 
	& = &  [D_0, [D_0, U]] = U_{00}.
\end{eqnarray}
Please note that $D_0 ( f(\Omega))=0$. Here, we carefully convert all the open derivatives into closed ones such that we can move them freely inside the trace. 
\begin{eqnarray}
	(Q^2)_{i} & = & [D_i, Q^2] =  Q[D_i,Q]+ [D_i,Q] Q = Q G_{i0} + G_{i0} Q \nonumber \\
	& =  & -Q E_{i} - E_{i} Q = -2Q E_{i} + [Q,E_i] =-2Q E_{i} + E_{i0} . \nonumber \\
	(Q^2)_{i} (Q^2)_{i} &=& 4 Q^2 E_i^2 -6 Q E_{i0} E_i -2 Q E_i E_{i0} + 2 E_{i00} E_i + E_{i0}^2.
\end{eqnarray}

\begin{eqnarray}
	(Q^2)_{ii} & = & [D_i,[D_i, Q^2]] = [D_i, Q[D_i,Q]+ [D_i,Q] Q]= [D_i,Q G_{i0} + G_{i0} Q] .\nonumber 
\end{eqnarray}
\begin{eqnarray}
	[D_i, Q G_{i0}] & =& Q[D_i,G_{i0}] + [D_i, Q] G_{i0} \nonumber \\
	& =& -Q E_{ii} + (G_{i0})^2. \nonumber 
\end{eqnarray}
\begin{eqnarray}
	[D_i,  G_{i0} Q] & =& - E_{ii} Q + (G_{i0})^2 = E_i^2 - Q E_{ii} + [Q, E_{ii}],
\end{eqnarray}
where, we have defined $E_i= G_{0i}=- G_{i0}$, $[D_i, Q] = [D_i,D_0]=G_{i0}$, and $[Q, E_{ii}]= E_{ii0}$.
Thus,  $(Q^2)_{ii}= [D_i,Q G_{i0}] + [D_i, G_{i0} Q]= - 2 QE_{ii} + 2 E_i^2 +E_{ii0}$.
\begin{eqnarray}
	\tilde{b}_2 &=& U^2 -\frac{1}{3} U_{ii}  + \frac{1}{6} (G_{ij})^2 + \frac{1}{3} (Q^2)_{ii}+ [Q^2, U] \nonumber \\
	& = & [U^2 -\frac{1}{3} U_{ii}  + \frac{1}{6} (G_{ij})^2] - \frac{1}{3} [2 QE_{ii} - 2 E_i^2 -E_{ii0}] + 2QU_0 -U_{00} \nonumber \\
	& = & [U^2 -\frac{1}{3} U_{ii}  + \frac{1}{6} (G_{ij})^2] + \frac{1}{3} [ 2 E_i^2 + E_{ii0} - 3 U_{00}] -\frac{2}{3} Q [E_{ii} -3 U_0]\nonumber \\
	& = & \tilde{b}_2^0 + \tilde{b}_{21}^T + Q \tilde{b}_{22}^T.
\end{eqnarray}

\begin{eqnarray}
	(Q^2)_{iijj} &= & [D_j,[D_j, - 2 QE_{ii} + 2 E_i^2 +E_{ii0}]]  \\
	& = & D_j[D_j(- 2 QE_{ii} + 2 E_i^2 +E_{ii0})] \nonumber \\
	&=& D_j[2E_j E_ii -2 Q E_{iij} + 2 E_i E_{ij} + E_{ii0j}+2E_{ij}E_i] \nonumber \\
	&=& 2 E_{ii} E_{jj} + 4 E_{ij} E_{ij} + 4 E_j E_{iij} + 2 E_i E_{ijj} +2 E_{ijj} E_i +E_{ii0jj} - 2 Q E_{iijj}. \nonumber
\end{eqnarray}

\begin{eqnarray}
	\tilde{b}_3   & =& U^3 - 2 [Q^2,U] U + U [ Q^2,U] + [Q^2, [Q^2,U]] - Q^2 U_{ii} + Q^2  (Q^2)_{ii} \nonumber \\ 
	& + &  \frac{1}{2} Q^2 (G_{ij})^2 + \frac{1}{2} (U_i -Q^2_i)^2 + \frac{1}{2} (U-Q^2) (G_{ij})^2  + \frac{1}{10} (J_i)^2 - \frac{1}{15} G_{ij} G_{jk} G_{ki}   \nonumber \\
	&=&  U^3  + \frac{1}{10} (J_i)^2 - \frac{1}{15} G_{ij} G_{jk} G_{ki} + \frac{1}{2} U (G_{ij})^2 + \frac{1}{2} (U_i)^2  + \frac{1}{10} U_{iijj} -\frac{1}{30} (J_{i})_{jji} \nonumber  \\
	& & + 4Q^2 U_{00} - 4Q U_{000}  -  U_{0000} \nonumber \\
	& &  + 2QUU_0-2(U_0)^2 -UU_{00} - 4QU_0 U +2U_{00} U   \\
	& & - Q^2 U_{ii} - 2 Q^3 E_{ii} + 2 Q^2  E_i^2 + Q^2 E_{ii0} \nonumber \\
	& & + \frac{1}{2} [4 Q^2 E_i^2 -6 Q E_{i0} E_i -2 Q E_i E_{i0} + 2 E_{i00} E_i + E_{i0}^2]\nonumber \\
	& & + \frac{1}{2} [ 2 Q U_i E_{i} + 2Q E_{i} U_i - 2U_{i0} E_{i}  - U_i E_{i0} - E_{i0} U_i] \nonumber \\
	& & + \frac{1}{10} [2 E_{ii} E_{jj} + 4 E_{ij} E_{ij} + 4 E_j E_{iij} + 2 E_i E_{ijj} +2 E_{ijj} E_i +E_{ii0jj} - 2 Q E_{iijj}]. \nonumber \\
	&=& (3!)\;\Big[\tilde{b}_3^{0} + \tilde{b}_{31}^{T} + Q\;\tilde{b}_{32}^{T} +  Q^2\;\tilde{b}_{33}^{T} + Q^3\;\tilde{b}_{34}^{T}\Big].
\end{eqnarray}

\subsection{ $ \mathbb{I}[k;l]$: Without Polyakov loop Contribution } \label{HKC-I}

Here, we must set $d=3-2\epsilon$ to recover the thermally corrected effective action for a four-dimensional theory. In the limit when we set the Polyakov loop contribution to zero, i.e. $\Omega \to 1$,  we find

\begin{eqnarray}
	\mathbb{I}[0;0] &=&  \frac{\mu^{2\epsilon}}{\beta (4\pi)^{d/2}} \Big(\frac{2\pi}{\beta}\Big)^{d}\;\;
	\Gamma (\frac{-d}{2})\Bigg[ |(\;m_{\beta})|^{+d} 
	+ 2\; \mathbb{CS} \Big(m_{\beta};\; \frac{-d}{2} \Big) \Bigg]  \nonumber \\
	&= & \frac{\mu^{2\epsilon}}{\beta (4\pi)^{(3-2\epsilon)/2}} \Big(\frac{2\pi}{\beta}\Big)^{(3-2\epsilon)}\;\;
	\Gamma (\frac{-3+2\epsilon}{2}) \Bigg[ (m_{\beta})^{+(3-2\epsilon)} 
	+ 2\;  \mathbb{CS} \Big(m_{\beta};\; \frac{-3+2\epsilon}{2} \Big) \Bigg] \nonumber \\
	&=&   \frac{\mu^{2\epsilon}}{\beta (4\pi)^{(3-2\epsilon)/2}} \Big(\frac{2\pi}{\beta}\Big)^{(3-2\epsilon)}\;\;
	\Gamma (\frac{-3+2\epsilon}{2}) \Bigg[ (m_{\beta})^{+(3-2\epsilon)} 
	+ 2\; \Big\{ 
	-\frac{(m_{\beta})^{3-2\epsilon}}{2} \nonumber \\
	&+& \frac{\sqrt{\pi} \Gamma(-2+\epsilon)}{2 \Gamma(-3/2+\epsilon)} (m_\beta)^{4-2\epsilon} 
	+ \frac{2 (m_\beta)^{2-\epsilon}}{(\sqrt{\pi})^{3-2\epsilon} \Gamma(-3/2+\epsilon)} \sum_{n=1}^{\infty} \frac{1}{n^{2-\epsilon}}  \mathbb{K}_{-2+\epsilon } (2\pi n m_\beta)
	\Big\}
	\Bigg] \nonumber \\
	&=&  \frac{m^3}{12\pi \beta} -  \frac{m^3}{12\pi \beta} + 2\; \Bigg\{ \frac{m^2 }{2\pi^2 \beta^2}    \Big(\frac{2\pi \mu^2 \beta}{m}\Big)^{\epsilon}   \sum_{n=1}^{\infty} \frac{1}{n^2} 
	\mathbb{K}_{-2} (n m \beta) + \frac{m^4}{32\pi^2} \Gamma(-2+\epsilon) \Big(\frac{2\pi \mu^2}{m^2}\Big)^{\epsilon} \Bigg\}. \nonumber
\end{eqnarray}

\begin{eqnarray}
	\mathbb{I}[0;1] &=&  \frac{\mu^{2\epsilon}}{\beta (4\pi)^{d/2}} \Big(\frac{2\pi}{\beta}\Big)^{d-2}\;\;
	\Gamma (\frac{2-d}{2}) \Bigg[ (\;m_{\beta})^{-2+d} 
	+ 2\; \mathbb{CS} \Big(m_{\beta};\; \frac{2-d}{2} \Big) \Bigg] \nonumber  \\
	& = &  \frac{\mu^{2\epsilon}}{\beta (4\pi)^{(3-2\epsilon)/2}} \Big(\frac{2\pi}{\beta}\Big)^{1-2\epsilon}\;\;
	\Gamma (\frac{-1+2\epsilon}{2}) \Bigg[ (\;m_{\beta})^{1-2\epsilon} 
	+ 2\; \mathbb{CS} \Big(m_{\beta};\; \frac{-1+2\epsilon}{2} \Big) \Bigg] \nonumber \\
	&=&  \frac{\mu^{2\epsilon}}{\beta (4\pi)^{(3-2\epsilon)/2}} \Big(\frac{2\pi}{\beta}\Big)^{1-2\epsilon}\;\; \Gamma (\frac{-1+2\epsilon}{2})  \Bigg[ (\;m_{\beta})^{1-2\epsilon} + 2\; \Big\{ -\frac{(m_{\beta})^{1-2\epsilon}}{2} \nonumber \\
	&+& \frac{\sqrt{\pi} \Gamma(-1+\epsilon)}{2 \Gamma(-1/2+\epsilon)} (m_\beta)^{2-2\epsilon} 
	+ \frac{2 m_\beta}{\sqrt{\pi} \Gamma(-1/2+\epsilon)} \sum_{n=1}^{\infty} \frac{1}{n}  \mathbb{K}_{-1} (2\pi n m_\beta)
	\Big\}
	\Bigg] \nonumber \\
	&=& -\frac{m}{8\pi \beta} + \frac{m}{8\pi \beta} + 2\; \Bigg\{\frac{m}{4\pi^2 \beta} \sum_{n=1}^{\infty} \frac{1}{n} 
	\mathbb{K}_{-1} (n m \beta) + \frac{m^2}{16\pi^2} \Gamma(-1+\epsilon) \Big(\frac{4\pi \mu^2}{m^2}\Big)^{\epsilon} \Bigg\}. \nonumber
\end{eqnarray}

\begin{eqnarray}
	\mathbb{I}[0;2] &=&  \frac{\mu^{2\epsilon}}{\beta (4\pi)^{d/2}} \Big(\frac{2\pi}{\beta}\Big)^{d-4}\;\;
	\Gamma (\frac{4-d}{2}) \Bigg[ (\;m_{\beta})^{-4+d} 
	+ 2\; \mathbb{CS} \Big(m_{\beta};\; \frac{4-d}{2} \Big) \Bigg] \nonumber  \\
	& = &  \frac{\mu^{2\epsilon}}{\beta (4\pi)^{(3-2\epsilon)/2}} \Big(\frac{2\pi}{\beta}\Big)^{-1-2\epsilon}\;\;
	\Gamma (\frac{1+2\epsilon}{2}) \Bigg[ (\;m_{\beta})^{-1-2\epsilon} 
	+ 2\; \mathbb{CS} \Big(m_{\beta};\; \frac{1+2\epsilon}{2} \Big) \Bigg] \nonumber \\
	&=&  \frac{\mu^{2\epsilon}}{\beta (4\pi)^{(3-2\epsilon)/2}} \Big(\frac{2\pi}{\beta}\Big)^{-1-2\epsilon}\;\; \Gamma (\frac{1+2\epsilon}{2})  \Bigg[ (\;m_{\beta})^{-1-2\epsilon} + 2\; \Big\{ -\frac{(m_{\beta})^{-1-2\epsilon}}{2} \nonumber \\
	&+& \frac{\sqrt{\pi} \Gamma(\epsilon)}{2 \Gamma(1/2+\epsilon)} (m_\beta)^{-2\epsilon} 
	+ \frac{2 \sqrt{\pi}}{ \Gamma(1/2+\epsilon)} \sum_{n=1}^{\infty} \frac{1}{n}  \mathbb{K}_{0} (2\pi n m_\beta)
	\Big\}
	\Bigg] \nonumber \\
	&=&  2\; \Bigg\{\frac{1}{8\pi^2 } \sum_{n=1}^{\infty}  
	\mathbb{K}_{0} (n m \beta) + \frac{1}{32\pi^2} \Gamma(\epsilon) \Big(\frac{4\pi \mu^2}{m^2}\Big)^{\epsilon} \Bigg\}. \nonumber
\end{eqnarray}

\begin{eqnarray}
	\mathbb{I}[0;3] &=&  \frac{\mu^{2\epsilon}}{\beta (4\pi)^{d/2}} \Big(\frac{2\pi}{\beta}\Big)^{d-6}\;\;
	\Gamma (\frac{6-d}{2}) \Bigg[ (\;m_{\beta})^{-6+d} 
	+ 2\; \mathbb{CS} \Big(m_{\beta};\; \frac{6-d}{2} \Big) \Bigg] \nonumber  \\
	&= &  \frac{\mu^{2\epsilon}}{\beta (4\pi)^{(3-2\epsilon)/2}} \Big(\frac{2\pi}{\beta}\Big)^{-3-2\epsilon}\;\;
	\Gamma (\frac{3+2\epsilon}{2}) \Bigg[ (\;m_{\beta})^{-3-2\epsilon} 
	+ 2\; \mathbb{CS} \Big(m_{\beta};\; \frac{3+2\epsilon}{2} \Big) \Bigg] \nonumber \\
	&=&  \frac{\mu^{2\epsilon}}{\beta (4\pi)^{(3-2\epsilon)/2}} \Big(\frac{2\pi}{\beta}\Big)^{-3-2\epsilon}\;\; \Gamma (\frac{3+2\epsilon}{2})  \Bigg[ (\;m_{\beta})^{-3-2\epsilon} + 2\; \Big\{ -\frac{(m_{\beta})^{-3-2\epsilon}}{2} \nonumber \\
	&+& \frac{\sqrt{\pi} \Gamma(1+\epsilon)}{2 \Gamma(3/2+\epsilon)} (m_\beta)^{-2-2\epsilon} 
	+ \frac{2 \sqrt{\pi}^{3+2\epsilon}\; (m_\beta)^{-1-\epsilon}}{ \Gamma(3/2+\epsilon)} \sum_{n=1}^{\infty} n  \mathbb{K}_{1} (2\pi n m_\beta)
	\Big\}
	\Bigg] \nonumber \\
	&=&  2\; \Bigg\{ \frac{1}{32 \pi^2 m^2}+ \frac{1}{16 \pi^2 m T }   \sum_{n=1}^{\infty} n\, \mathbb{K}_{1} (n m \beta) \Bigg\}. \nonumber
\end{eqnarray}

Note that here, $m_\beta = (m_\beta^2)^{1/2}$.

\subsection{Sum of Modified Bessel Function}\label{app:sum_bessel}

We have used the method given in Ref.~\cite{Paris} to compute the following sum of the Bessel's functions employing suitable Mellin transformation and identifying the associated poles of different orders. 

Let us consider two functions $F(s)$ and $f(x)$ that are related to each other through Mellin transformation (MT) as:
\begin{eqnarray}\label{app:MT}
	F(s) &=& \int_{0}^{\infty} f(x) \; x^{s-1}\; dx \nn \\
	f(x) &=& \frac{1}{2\pi \iu} \int_{\sigma-\iu \infty}^{\sigma+\iu \infty} F(s)\; x^{-s}\; ds,
\end{eqnarray}
where $\sigma=Re(s)$ and $\sigma_1 < \sigma < \sigma_2$. Here, we will introduce some properties that will be used latter :
\begin{eqnarray}\label{app:properties}
(i)\;\; g(n+a) &=&   \frac{1}{2\pi \iu} \int_{\sigma-\iu \infty}^{\sigma+\iu \infty} F(s)\; (x+a)^{-s}\; ds, \nn \\
		\sum_{n=0}^{\infty} g(n+a) &=&   \frac{1}{2\pi \iu} \int_{\sigma-\iu \infty}^{\sigma+\iu \infty} F(s)\;	\sum_{n=0}^{\infty} (x+a)^{-s}\; ds \nn \\
		&=&  \frac{1}{2\pi \iu} \int_{\sigma-\iu \infty}^{\sigma+\iu \infty} F(s)\;	 \zeta(s;a)\; ds, \\
(ii)\;\;		\lim_{s \to 1} \Big[ \zeta(s;a) - \frac{1}{s-1}\Big] & = &  -\frac{\Gamma^{'}(a)}{\Gamma(a)} = -\psi (a), \\
(iii)\;\;	 Res \Big[\Gamma(x)\Big]\Big|_{x=-n} &=& \frac{(-1)^n}{\Gamma(n+1)}, \\
(iv)\;\; \Gamma(s) \Gamma(1-s) &=& \frac{\pi}{\sin(\pi s)}, \\
(v) \;\;  \int_{0}^{\infty} \mathbb{K}_{\nu}(x) \; x^{s-1}\; dx &=& 2^{s-2} \Gamma(s/2-\nu/2) \Gamma(s/2+\nu/2).
\end{eqnarray}

We aim to compute the following sum:
\begin{equation}
    S_{\alpha,\nu} (z) = \sum_{n=1}^{\infty} \frac{1}{n^\alpha} \mathbb{K}_{\nu}(nz),
\end{equation}
where, $arg|z| < \pi/2$, and $\nu \geq 0$\footnote{Since  $\mathbb{K}_{\nu}(z)= \mathbb{K}_{-\nu}(z)$.}.
Let us define, $ \frac{1}{n^\alpha} \mathbb{K}_{\nu}(nz)=z^\alpha \; \frac{\mathbb{K}_{\nu}(nz)}{(nz)}=f(x)$, where $x=nz$.
Employing \ref{app:properties}, we find
\begin{eqnarray}
	 \int_{0}^{\infty} \frac{\mathbb{K}_{\nu}(x)}{x^\alpha} \; x^{s-1}\; dx &=& 2^{-\alpha+s-2} \Gamma(s/2-\alpha/2-\nu/2) \Gamma(s/2-\alpha/2+\nu/2).
\end{eqnarray}	
Performing inverse MT, \ref{app:MT}, we find
\begin{eqnarray}
	\frac{\mathbb{K}_{\nu}(x)}{x^\alpha} &=& \frac{1}{8\pi \iu} \int_{\sigma-\iu \infty}^{\sigma+\iu \infty} 2^{-\alpha+s} \Gamma(s/2-\alpha/2-\nu/2) \Gamma(s/2-\alpha/2+\nu/2)\; x^{-s}\; ds. 
\end{eqnarray}
Thus,
\begin{eqnarray}\label{app:besselsum}
\sum_{n=1}^{\infty}	\frac{\mathbb{K}_{\nu}(nz)}{n^\alpha} &=& \frac{1}{8\pi \iu} \int_{\sigma-\iu \infty}^{\sigma+\iu \infty} \Big[\frac{z}{2}\Big]^{\alpha-s} \Gamma(s/2-\alpha/2-\nu/2) \Gamma(s/2-\alpha/2+\nu/2)\; \Big[ \sum_{n=1}^{\infty}n^{-s}\Big]\; ds. \nn \\
&=& \frac{1}{8\pi \iu} \int_{\sigma-\iu \infty}^{\sigma+\iu \infty} \Big[\frac{z}{2}\Big]^{\alpha-s} \Gamma(s/2-\alpha/2-\nu/2) \Gamma(s/2-\alpha/2+\nu/2)\; \zeta(s)\; ds.
\end{eqnarray}
Here, $|arg(z)| <\pi/2; c>max\{1,\alpha \pm \nu\}$.
It is evident from this form that the sum can be written as the sum of residues of the integrand $\mathbb{I}(s)=\frac{1}{4}\Big[\frac{z}{2}\Big]^{\alpha-s} \Gamma(s/2-\alpha/2-\nu/2) \Gamma(s/2-\alpha/2+\nu/2)\; \zeta(s)$,  \ref{app:besselsum}, and sources of the poles are as follows:
\begin{itemize}
	\item at $s=1$, $\zeta(s)$ possesses simple pole. 
	\item  infinite sequences of simple poles at $s_r^+= \alpha+\nu-2r$;\;\; $\forall \;\; r=0,1,2,\cdots$.
		\item  infinite sequences of double poles at $s_r^+= \alpha-\nu-2r$;\;\; $\forall \;\; r=0,1,2,\cdots$.
\end{itemize}
where we consider the case with $\nu=N$.
The residue at $s=1$ is computed by expanding $\mathbb{I}$ around $s=1+\epsilon$, and identifying coefficients of $1/\epsilon$ that reads as
\begin{eqnarray}\label{app:res1}
	Res(s=1)&=& \frac{1}{4}\Big[\frac{z}{2}\Big]^{\alpha-1} \Gamma(1/2-\alpha/2-N/2) \Gamma(1/2-\alpha/2+N/2).
\end{eqnarray}

The residue at $s=s_r^+$ is computed by expanding $\mathbb{I}$ around $s=s_r^{+} +\epsilon$, and identifying coefficients of $1/\epsilon$ that reads as
\begin{eqnarray}\label{app:res2}
	\mathbb{I}(s=s_r^+ + \epsilon)& = & \frac{1}{4}\Big[\frac{z}{2}\Big]^{2r-N-\epsilon} \Gamma(-r+\epsilon/2) \Gamma(N-r+\epsilon/2) \zeta(\alpha+N-2r+\epsilon), \nn \\
		Res(s=s_r^+) &=&  \frac{1}{2} \sum_{r=0}^{N-1} \frac{(-1)^{r}}{r!} \Gamma(N-r) \zeta (s_r^+)  \Big[\frac{z}{2}\Big]^{2r-N}. \nonumber 
\end{eqnarray}

The residue at $s=s_r^-$ is computed by expanding $\mathbb{I}$ around $s=s_r^{-} +\epsilon$, and identifying coefficients of $1/\epsilon^2$ that reads as
\begin{eqnarray}\label{app:res2}
	\mathbb{I}(s=s_r^{-} + \epsilon)& = & \frac{1}{4}\Big[\frac{z}{2}\Big]^{2r+N-\epsilon} \Gamma(-r+\epsilon/2) \Gamma(-N-r+\epsilon/2) \zeta(\alpha-N-2r+\epsilon), \nn \\
	Res(s=s_r^-) &=&  \sum_{r=0}^{\infty} \frac{\zeta (s_r^-) }{r!\; (r+N)!} \Big[\frac{-z}{2}\Big]^{2r+N} \Big\{ \frac{\zeta^{'} (s_r^-)}{\zeta (s_r^-)} + \frac{1}{2} \psi(r+1) + \frac{1}{2} \psi (r+N+1) - ln (z/2)\Big \}. \nonumber 
\end{eqnarray}

Taking care of all such contributions we can express the sum as  \cite{Paris}
\begin{eqnarray}
   S_{\alpha,N} (z) &=& \frac{1}{4} \Big[ \frac{z}{2}\Big]^{\alpha-1} \Gamma\Bigg[\frac{1-\alpha + N}{2} \Big]  
   \Gamma\Bigg[\frac{1-\alpha - N}{2} \Big] \nonumber \\
&+& \frac{1}{2} \sum_{r=0}^{N-1} \frac{(-1)^{r}}{r!} \Gamma(N-r) \zeta (s_r^+)  \Big[\frac{z}{2}\Big]^{2r-N} \nonumber \\
&+& \sum_{r=0}^{\infty} \frac{\zeta (s_r^-) }{r!\; (r+N)!} \Big[\frac{-z}{2}\Big]^{2r+N} \Big\{ \frac{\zeta^{'} (s_r^-)}{\zeta (s_r^-)} + \frac{1}{2} \psi(r+1) + \frac{1}{2} \psi (r+N+1) - ln (z/2)\Big \} . \nonumber
\end{eqnarray}
 Here, poles occur at $s_r^+= \alpha +N-2r$, and $s_r^-= \alpha - N-2r$. The di-gamma function $\psi(q+1)$ is defined as $\psi(q+1)= -\gamma + \sum_{n=1}^{\infty} \frac{q}{(n+q)n}=-\gamma + \sum_{n=1}^{q} \frac{1}{n}$.
 For example,
 \begin{eqnarray}
     \psi(1) & = & - \gamma;\;\;\;\;  \psi(3)  =  - \gamma + \frac{3}{2}.
 \end{eqnarray}

 Let us define three terms as follows \cite{Paris}
 \begin{eqnarray}
   S_1 &=& \frac{1}{4} \Big[ \frac{z}{2}\Big]^{\alpha-1} \Gamma\Bigg[\frac{1-\alpha + N}{2} \Big]  
   \Gamma\Bigg[\frac{1-\alpha - N}{2} \Big] ; \nn \\
S_2 &=& \frac{1}{2} \sum_{r=0}^{N-1} \frac{(-1)^{r}}{r!} \Gamma(N-r) \zeta (s_r^+)  \Big[\frac{z}{2}\Big]^{2r-N};\\
S_3 &=& \sum_{r=0}^{\infty} \frac{\zeta (s_r^-) }{m!\; (r+N)!} \Big[\frac{-z}{2}\Big]^{2r+N} \Big\{ \frac{\zeta^{'} (s_r^-)}{\zeta (s_r^-)} + \frac{1}{2} \psi(r+1) + \frac{1}{2} \psi (r+N+1) - ln (z/2)\Big \} .\nn
\end{eqnarray}

Using this method we compute some of the Bessel function sums  which will be useful in the computation of the leading order thermal heat coefficients.

\subsection*{\fbox{$\mathbb{S}[0;0]$}}
We want to compute the following sum
\begin{equation}
  \mathbb{S}[0;0]  = \frac{m^2 }{\pi^2 \beta^2}   \Big(\frac{2\pi \mu^2 \beta}{m}\Big)^{\epsilon}     \sum_{n=1}^{\infty} \frac{1}{n^2} 
  \mathbb{K}_{-2} (n m \beta).
\end{equation}

We define 

    \begin{equation}
    S_{\alpha,\nu} (nz) = \sum_{n=1}^{\infty} \frac{1}{n^2} 
  \mathbb{K}_{-2} (n m \beta).
\end{equation}
Thus, we have $\alpha=2,N=2, z=m \beta$. Here, poles occur at $s_r^+= \alpha +N-2r=4-2r$ for $r=0,1$, and $s_r^-= \alpha - N-2r=-2r$ for $r\geq 1$.

The residue for simple pole at $s=1$ is given as
\begin{equation}
       S_1 = \frac{1}{4} \Big[ \frac{z}{2}\Big]^{\alpha-1} \Gamma\Bigg[\frac{1-\alpha + N}{2} \Big]  
   \Gamma\Bigg[\frac{1-\alpha - N}{2} \Big] = \frac{\pi m\beta}{6}.
\end{equation}

Residue at $s=s_r^+$ is given as 
\begin{eqnarray}
S_2 & = & \frac{1}{2} \sum_{r=0}^{N-1} \frac{(-1)^{r}}{r!} \Gamma(N-r) \zeta (s_r^+)  \Big[\frac{z}{2}\Big]^{2r-N} \nonumber \\
&=& \frac{\pi^4}{45\; m^2 \beta^2} -\frac{\pi^2}{12} .
\end{eqnarray}

Residue at $s=s_r^-$ is given as 
\begin{eqnarray}
    S_3 &=& \sum_{r=0}^{\infty} \frac{\zeta (s_r^-) }{r!\; (r+N)!} \Big[\frac{-z}{2}\Big]^{2r+N} \Big\{ \frac{\zeta^{'} (s_r^-)}{\zeta (s_r^-)} + \frac{1}{2} \psi(r+1) + \frac{1}{2} \psi (r+N+1) - ln (z/2)\Big \} \nonumber \\
    &=& \sum_{r=0}^{\infty} \frac{\zeta (-2r) }{r!\; (r+N)!} \Big[\frac{-z}{2}\Big]^{2r+N} \Big\{ \frac{\zeta^{'} (-2r)}{\zeta (-2r)} + \frac{1}{2} \psi(r+1) + \frac{1}{2} \psi (r+N+1) - ln (z/2)\Big \} \nonumber \\
    &=& \frac{\zeta (0) }{0!\; (2)!} \Big[\frac{-z}{2}\Big]^{2} \Big\{ \frac{\zeta^{'} (0)}{\zeta (0)} + \frac{1}{2} \psi(1) + \frac{1}{2} \psi (3) - ln (z/2)\Big \} \nonumber \\
 &+&  \sum_{r=1}^{\infty}  \frac{\zeta (-2r) }{r!\; (r+2)!} \Big[\frac{-z}{2}\Big]^{2r+2} \Big\{ \frac{\zeta^{'} (-2r)}{\zeta (-2r)} + \frac{1}{2} \psi(r+1) + \frac{1}{2} \psi (r+3) - ln (z/2)\Big \} \nonumber \\
 &=& \frac{m^2 \beta^2}{16} \Big\{ ln (\frac{z}{4\pi}) +\gamma -\frac{3}{4} \Big \} 
 + \sum_{r=1}^{\infty} \frac{ \zeta^{'} (-2r) \Big[\frac{-z}{2}\Big]^{2r+2} } {r!\; (r+2)!} \nonumber \\
  &=& \frac{m^2 \beta^2}{16} \Big\{ ln (\frac{m\beta}{4\pi}) +\gamma -\frac{3}{4} \Big \} 
  - \frac{\zeta(3)\; m^4\; \beta^4}{3\; (2)^7\; \pi^2} + \cdots
\end{eqnarray}

We use the following relation \cite{Paris}
\begin{equation}
     \zeta^{'} (-2r) = \frac{1}{2} \Big[\frac{-1}{(2\pi)^2}\Big]^r \zeta(2r+1) \Gamma(2r+1). 
\end{equation}

Final contribution reads as
\begin{eqnarray}\label{S00}
      \mathbb{S}[0;0]  &= & \frac{m^2 }{\pi^2 \beta^2}   \Big(\frac{2\pi \mu^2 \beta}{m}\Big)^{\epsilon}    \sum_{n=1}^{\infty} \frac{1}{n^2} 
  \mathbb{K}_{-2} (n m \beta) 
  = \frac{m^2 }{\pi^2 \beta^2} \Big[ S_1 \; + S_2\; + \; S_3 \Big]  \nonumber \\
  &=& \frac{m^2 }{\pi^2 \beta^2} \Big[ \frac{\pi m\beta}{6} \; +\; 
  \Big\{ \frac{\pi^4}{45\; m^2 \beta^2} -\frac{\pi^2}{12}\Big\}  \; +\;
  \Big\{   \frac{m^2 \beta^2}{16} \Big\{ ln (\frac{m\beta}{4\pi}) +\gamma -\frac{3}{4} \Big \} 
  - \frac{\zeta(3)\; m^4\; \beta^4}{3\; (2)^7\; \pi^2} + \cdots   \Big \}
  \Big] \nonumber \\
  &=&  \frac{m^3}{6 \pi \beta} \; +\; \frac{\pi^2}{45\; \beta^4} - \frac{ m^2}{12\; \beta^2} +\frac{m^4}{16 \pi^2}
  \Big\{ ln\Big[\frac{m\beta e^\gamma}{4\pi} \Big] -\frac{3}{4}\Big \} -\frac{2\zeta(3) \; m^6 \; \beta^2}{3 \; (4\pi)^4} +\cdots \nonumber 
\end{eqnarray}

\subsection*{\fbox{$\mathbb{S}[0;1]$}}
We want to compute the following sum

\begin{equation}
  \mathbb{S}[0;1]  = \frac{m}{2\pi^2 \beta}  \sum_{n=1}^{\infty} \frac{1}{n} 
  \mathbb{K}_{-1} (n m \beta).
\end{equation}

We define 
    \begin{equation}
    S_{\alpha,\nu} (nz) = \sum_{n=1}^{\infty} \frac{1}{n} 
  \mathbb{K}_{-1} (n m \beta).
\end{equation}
Thus, we have $\alpha=1,N=1, z=m \beta$. Here, poles occur at $s_r^+= \alpha +N-2r=2-2r$ for $r=0$, and $s_r^-= \alpha - N-2r=-2r$ for $r\geq 1$. Following the similar footsteps described in the previous case, the total residue is computed in terms of $S_1,S_2,S_3$ that are computed below. 

\begin{equation}
       S_1 = \frac{1}{4} \Big[ \frac{z}{2}\Big]^{\alpha-1} \Gamma\Bigg[\frac{1-\alpha + N}{2} \Big]  
   \Gamma\Bigg[\frac{1-\alpha - N}{2} \Big] = -\frac{\pi}{ 2}.
\end{equation}

\begin{eqnarray}
S_2 & = & \frac{1}{2} \sum_{r=0}^{N-1} \frac{(-1)^{r}}{r!} \Gamma(N-r) \zeta (s_r^+)  \Big[\frac{z}{2}\Big]^{2r-N} 
= \frac{\pi^2}{6m\beta}. 
\end{eqnarray}

\begin{eqnarray}
    S_3 &=& \sum_{r=0}^{\infty} \frac{\zeta (s_r^-) }{r!\; (r+N)!} \Big[\frac{-z}{2}\Big]^{2r+N} \Big\{ \frac{\zeta^{'} (s_r^-)}{\zeta (s_r^-)} + \frac{1}{2} \psi(r+1) + \frac{1}{2} \psi (r+N+1) -ln (z/2)\Big \} \nonumber \\
    &=& \sum_{r=0}^{\infty} \frac{\zeta (-2r) }{r!\; (r+N)!} \Big[\frac{-z}{2}\Big]^{2r+N} \Big\{ \frac{\zeta^{'} (-2r)}{\zeta (-2r)} + \frac{1}{2} \psi(r+1) + \frac{1}{2} \psi (r+N+1) - ln (z/2)\Big \} \nonumber \\
    &=& \frac{\zeta (0) }{ (1)!} \Big[\frac{-z}{2}\Big]^{1} \Big\{ \frac{\zeta^{'} (0)}{\zeta (0)} + \frac{1}{2} \psi(1) + \frac{1}{2} \psi (2) -ln (z/2)\Big \} \nonumber \\
 &+&  \sum_{r=1}^{\infty}  \frac{\zeta (-2r) }{r!\; (r+1)!} \Big[\frac{-z}{2}\Big]^{2r+1} \Big\{ \frac{\zeta^{'} (-2r)}{\zeta (-2r)} + \frac{1}{2} \psi(r+1) + \frac{1}{2} \psi (r+2) -ln (z/2)\Big \} \nonumber \\
 &=& -\frac{m\beta}{4} \Big\{ ln\Big[ \frac{m\beta e^\gamma}{4\pi}\Big] -\frac{1}{2} \Big \} 
 + \sum_{r=1}^{\infty} \frac{(-1)^r}{2} \frac{\zeta(2r+1) \Gamma(2r+1)}{r!\; (r+1)! (2\pi)^{2r}} \Big[\frac{-z}{2}\Big]^{2r+1} \nonumber \\
 &=& -\frac{m\beta}{4} \Big\{ ln\Big[ \frac{m\beta e^\gamma}{4\pi}\Big] -\frac{1}{2} \Big \} 
 + \frac{\zeta(3)}{2^4\; (2\pi)^2} m^3 \beta^3+\cdots .
 \end{eqnarray}

Final contribution reads as
    \begin{eqnarray}
  \mathbb{S}[0;1] & =  & \frac{m}{2\pi^2 \beta} \Bigg[ -\frac{\pi}{ 2} + \frac{\pi^2}{6m\beta} -\frac{m\beta}{4} \Big\{ ln\Big[ \frac{m\beta e^\gamma}{4\pi}\Big] -\frac{1}{2} \Big \} 
 + \frac{\zeta(3)}{2^4\; (2\pi)^2} m^3 \beta^3 \cdots   \Bigg] \nonumber \\
 & =& -\frac{m}{4\pi \beta} + \frac{1}{12\beta^2} -\frac{m^2}{8\pi^2} \Big\{ ln\Big[ \frac{m\beta e^\gamma}{4\pi}\Big] -\frac{1}{2}\Big \} +\frac{2m^4 \beta^2}{(4\pi)^4} \zeta(3) +\cdots
\end{eqnarray}

\subsection*{\fbox{ $\mathbb{S}[0;2]$}}

We define 

\begin{equation}
	S_{\alpha,\nu} (nz) =  \frac{1}{4\pi^2}  \sum_{n=1}^{\infty} 
	\mathbb{K}_{0} (n m \beta).
\end{equation}
Thus, we have $\alpha=0,N=0, z=m \beta$. Here, poles do not occur at $s_r^+= \alpha +N-2r=-2r$ for any $r$, but double poles at $s_r^-= \alpha - N-2r=-2r$ for $r\geq 1$. Here, the total relevant contribution of this sum can be written in terms of $S_1,S_2,S_3$, computed below.

\begin{equation}
	S_1 = \frac{1}{4} \Big[ \frac{z}{2}\Big]^{\alpha-1} \Gamma\Bigg[\frac{1-\alpha + N}{2} \Big]  
	\Gamma\Bigg[\frac{1-\alpha - N}{2} \Big] = \frac{\pi}{2 m \beta}.
\end{equation}

\begin{eqnarray}
	S_2 & = & \frac{1}{2} \sum_{r=0}^{N-1} \frac{(-1)^{r}}{r!} \Gamma(N-r) \zeta (s_r^+)  \Big[\frac{z}{2}\Big]^{2r-N} 
	=0 . \nonumber
\end{eqnarray}

\begin{eqnarray}
	S_3 &=& \sum_{r=0}^{\infty} \frac{\zeta (s_r^-) }{r!\; (r+N)!} \Big[\frac{-z}{2}\Big]^{2r+N} \Big\{ \frac{\zeta^{'} (s_r^-)}{\zeta (s_r^-)} + \frac{1}{2} \psi(r+1) + \frac{1}{2} \psi (r+N+1) -ln (z/2)\Big \} \nonumber \\
	&=& \frac{ 1}{2}  ln\Big[ \frac{z e^\gamma}{4\pi}\Big] +  \sum_{r=1}^{\infty} \frac{\zeta^{'} (-2r) }{r!\; r!} \Big[\frac{-z}{2}\Big]^{2r} \nonumber \\
	&=&  \frac{ 1}{2}  ln\Big[ \frac{z e^\gamma}{4\pi}\Big] +  \sum_{r=1}^{\infty} \frac{(-1)^r}{2} \frac{\zeta(2r+1) \Gamma(2r+1)}{r!\; r!\; (2\pi)^{2r}} \Big[\frac{z}{2}\Big]^{2r} \nonumber \\
	&=&  \frac{ 1}{2}  ln\Big[ \frac{m\beta\; e^\gamma}{4\pi}\Big] - \frac{\zeta(3) \Gamma(3)}{2 (2\pi)^2} \Big[ \frac{m\beta}{2} \Big]^2 +  \frac{\zeta(5) \Gamma(5)}{2 (2!)^2 (2\pi)^4} \Big[ \frac{m\beta}{2} \Big]^4 + \cdots
\end{eqnarray}

Final contribution reads as
\begin{eqnarray}
	\mathbb{S}[0;2] &=& \frac{1}{4\pi^2} \Big[    \frac{\pi}{2m \beta} + \Big\{  \frac{ 1}{2}  ln\Big[ \frac{m\beta\; e^\gamma}{4\pi}\Big] - \frac{\zeta(3) \Gamma(3)}{2 (2\pi)^2} \Big[ \frac{m\beta}{2} \Big]^2 +  \frac{\zeta(5) \Gamma(5)}{2 (2\pi)^4} \Big[ \frac{m\beta}{2} \Big]^4 + \cdots   \Big\}    \Big ]. \nn
\end{eqnarray}

\subsection*{\fbox{$\mathbb{S}[0;3]$}}
We want to compute the following sum
\begin{equation}
  \mathbb{S}[0;3]  =  \frac{\beta }{8\pi^2 m} \sum_{n=1}^{\infty} n\;
  \mathbb{K}_{-1} (n m \beta) .
\end{equation}

We define 

    \begin{equation}
    S_{\alpha,\nu} (nz) = \sum_{n=1}^{\infty} n\; 
  \mathbb{K}_{-1} (n m \beta).
\end{equation}
Thus, we have $\alpha=-1,N=1, z=m \beta$. Here, poles occur at $s_r^+= \alpha +N-2r=-2r$ for $r=0$, and $s_r^-= \alpha - N-2r=-2-2r$ for $r\geq 0$. Here, the total contributions cane be written as sum of residues computed in the form of functions $S_1,S_2,S_3$.

\begin{equation}
       S_1 = \frac{1}{4} \Big[ \frac{z}{2}\Big]^{\alpha-1} \Gamma\Bigg[\frac{1-\alpha + N}{2} \Big]  
   \Gamma\Bigg[\frac{1-\alpha - N}{2} \Big] = \frac{\pi}{ m^2 \beta^2}.
\end{equation}

\begin{eqnarray}
S_2 & = & \frac{1}{2} \sum_{r=0}^{N-1} \frac{(-1)^{r}}{r!} \Gamma(N-r) \zeta (s_r^+)  \Big[\frac{z}{2}\Big]^{2r-N} 
= -\frac{1}{2m\beta}. \nonumber
\end{eqnarray}

\begin{eqnarray}
    S_3 &=& \sum_{r=0}^{\infty} \frac{\zeta (s_r^-) }{r!\; (r+N)!} \Big[\frac{-z}{2}\Big]^{2r+N} \Big\{ \frac{\zeta^{'} (s_r^-)}{\zeta (s_r^-)} + \frac{1}{2} \psi(r+1) + \frac{1}{2} \psi (r+N+1) -ln (z/2)\Big \} \nonumber \\
    &=& \sum_{r=0}^{\infty} \frac{\zeta^{'} (-2-2r) }{r!\; (r+1)!} \Big[\frac{-z}{2}\Big]^{2r+1} \nonumber \\
     &=& \sum_{r=0}^{\infty} \frac{(-1)^{r+1}}{2} \frac{\zeta(2r+1) \Gamma(2r+3)}{r!\; (r+1)!\; (2\pi)^{2r+2}} \Big[-\frac{z}{2}\Big]^{2r+1} \nonumber \\
     &=& \frac{m\beta}{2 (2\pi)^2} \zeta(3) -\frac{3 m^3 \beta^3}{8(2\pi)^4} \zeta(5)+\cdots
\end{eqnarray}

Final contribution reads as
\begin{eqnarray}
  \mathbb{S}[0;3] & =&  \frac{\beta }{8\pi^2 m} \Bigg[\frac{\pi}{ m^2 \beta^2}  -\frac{1}{2m\beta} + \frac{m\beta}{2 (2\pi)^2} \zeta(3) -\frac{3 m^3 \beta^3}{8(2\pi)^4} \zeta(5)+\cdots    \Bigg] \nonumber \\
  &=& \frac{1}{8\pi m^3 \beta} + \frac{ \beta^2}{2 (2\pi)^2} \zeta(3) -\frac{12m^2 \beta^4}{(4\pi)^6 } \zeta(5)+\cdots
\end{eqnarray}

\subsection*{\fbox{$\hat{\mathbb{S}}_\Omega[0;0]$}}

\begin{equation}
 \hat{\mathbb{S}}_\Omega[0;0]  =  \sum_{l=0}^{\infty}  \sum_{n=1}^{\infty} n^{2l-2}\;
  \mathbb{K}_{-2} (n m \beta) .
\end{equation}
Thus, we have $\alpha=2-2l,N=2, z=m \beta$. Here, poles occur at $s_r^+= \alpha +N-2r=4-2l-2r$, and $s_r^-= \alpha - N-2r=-2l-2r$. Here, we have $l \geq 0$. Here, the non-zero residue is expressed in terms of $S_1,S_2,S_3$, computed below.

\begin{eqnarray}
       S_1 &=& \sum_{l=0}^{\infty}  \frac{1}{4} \Big[ \frac{z}{2}\Big]^{\alpha-1} \Gamma\Bigg[\frac{1-\alpha + N}{2} \Big]  
   \Gamma\Bigg[\frac{1-\alpha - N}{2} \Big] \nn \\
   & =& \sum_{l=0}^{\infty}  \frac{1}{4} \Big[ \frac{z}{2}\Big]^{1-2l} \Gamma\Bigg[\frac{1+2l}{2} \Big]  
   \Gamma\Bigg[\frac{-3-2l}{2} \Big].
\end{eqnarray}

\begin{eqnarray}
S_2 & = & \sum_{l=0}^{\infty}  \frac{1}{2} \sum_{r=0}^{N-1} \frac{(-1)^{r}}{r!} \Gamma(N-r) \zeta (s_r^+)  \Big[\frac{z}{2}\Big]^{2r-N} \nn \\
&=& \sum_{l=0}^{\infty}  \frac{1}{2} \Big [ \zeta (4-2l) \Big[\frac{z}{2}\Big]^{-2} - \zeta (2-2l)  \Big].
\end{eqnarray}

\begin{eqnarray}
    S_3 &=& \sum_{l=0}^{\infty}  \sum_{r=0}^{\infty} \frac{\zeta (s_r^-) }{r!\; (r+N)!} \Big[\frac{-z}{2}\Big]^{2r+N} \Big\{ \frac{\zeta^{'} (s_r^-)}{\zeta (s_r^-)} + \frac{1}{2} \psi(r+1) + \frac{1}{2} \psi (r+N+1) -ln (z/2)\Big \} \nn \\
    &=& \sum_{l=0}^{\infty}  \sum_{r=0}^{\infty} \frac{1}{r!\; (r+2)!} \Big[\frac{-z}{2}\Big]^{2r+2} \Big\{ 
    \zeta^{'}(-2l-2r) \nn \\
    & & + \zeta (-2l-2r) \Big[ \frac{1}{2} \psi(r+1) + \frac{1}{2} \psi(r+3) -ln(m\beta/2) \Big]\Big\} \nn \\
    &=& \sum_{l=0}^{\infty}  \Bigg[ \sum_{r=0}^{\infty} \frac{1}{r!\; (r+2)!} \Big[-\frac{m\beta}{2}\Big]^{2r+2} \Big\{ \frac{1}{2} \Big[-\frac{1}{(2\pi)^2} \Big]^{r+l} \zeta(2r+2l+1) \Gamma(2r+2l+1)\Big\} \nn \\
    & + &   \frac{1}{2} \Big[-\frac{m\beta}{2}\Big]^{2} \zeta(0) \Big[ -\gamma+3/4 -ln(m\beta/2) \Big] \Bigg].
\end{eqnarray}

Finally, collecting all the contributions, the sum reads as
\bea
 \hat{\mathbb{S}}_\Omega[0;0] &=&   \sum_{l=0}^{\infty}  \frac{1}{4} \Big[ \frac{m\beta}{2}\Big]^{1-2l} \Gamma\Bigg[\frac{1+2l}{2} \Big]     \Gamma\Bigg[\frac{-3-2l}{2} \Big] \nn \\
 &+& \sum_{l=0}^{\infty}  \frac{1}{2} \Big [ \zeta (4-2l) \Big[\frac{m\beta}{2}\Big]^{-2} - \zeta (2-2l)  \Big] \nn \\
  &+& \sum_{l=0}^{\infty}  \Bigg[ \sum_{r=0}^{\infty} \frac{1}{r!\; (r+2)!} \Big[-\frac{m\beta}{2}\Big]^{2r+2} \Big\{ \frac{1}{2} \Big[-\frac{1}{(2\pi)^2} \Big]^{r+l} \zeta(2r+2l+1) \Gamma(2r+2l+1)\Big\} \nn \\
    & + &   \frac{1}{2} \Big[-\frac{m\beta}{2}\Big]^{2} \zeta(0) \Big[ -\gamma+3/4 -ln(m\beta/2) \Big] \Bigg].
\eea

\subsection{ Relation between $\mathbb{I}_F[k;l]$ and $\mathbb{I}_B[k;l]$}\label{HKC-F}
\begin{eqnarray}
    \mathbb{I}_F[0;0] (\beta)  &=& \sum_{n} \frac{\mu^{2\epsilon}}{\beta (4\pi)^{d/2}} 
     \Big[ m^2+ \Big(\frac{2\pi}{\beta}\Big)^2 ((n+1/2)+\tilde{n})^2\Big]^{\frac{d}{2}}
      \Gamma (\frac{-d}{2}) \nonumber \\
      &=&     \sum_{p=-\infty}^{\infty} \frac{\mu^{2\epsilon}}{\beta (4\pi)^{d/2}} \Big[\frac{2\pi}{2\beta}\Big]^{d}
     \Big[ (2m_\beta)^2+ (p)^2 \Big]^{\frac{d}{2}}
      \Gamma (\frac{-d}{2}) \nonumber \\
      &=& 2\;  \sum_{p=1}^{\infty} \frac{\mu^{2\epsilon}}{\beta (4\pi)^{d/2}} \Big[\frac{2\pi}{2\beta}\Big]^{d}
     \Big[ (2m_\beta)^2+ (p)^2 \Big]^{\frac{d}{2}}
      \Gamma (\frac{-d}{2}) \nonumber \\
      &=& 2\;  \sum_{r=1}^{\infty} \frac{\mu^{2\epsilon}}{\beta (4\pi)^{d/2}} \Big[\frac{2\pi}{2\beta}\Big]^{d}
     \Big[ (2m_\beta)^2+ (r)^2 \Big]^{\frac{d}{2}}
      \Gamma (\frac{-d}{2}) \nonumber \\
      & & - 2\;  \sum_{l=1}^{\infty} \frac{\mu^{2\epsilon}}{\beta (4\pi)^{d/2}} \Big[\frac{2\pi}{2\beta}\Big]^{d}
     \Big[ (2m_\beta)^2+ (2l)^2 \Big]^{\frac{d}{2}}
      \Gamma (\frac{-d}{2}) \nonumber \\
       &=& 4\;  \sum_{r=1}^{\infty} \frac{\mu^{2\epsilon}}{(2\beta) (4\pi)^{d/2}} \Big[\frac{2\pi}{(2\beta)}\Big]^{d}
     \Big[ \Big\{\frac{m(2\beta)}{2\pi}\Big\}^2+ (r)^2 \Big]^{\frac{d}{2}}
      \Gamma (\frac{-d}{2}) \nonumber \\
      & & - 2\;  \sum_{l=1}^{\infty} \frac{\mu^{2\epsilon}}{\beta (4\pi)^{d/2}} \Big[\frac{2\pi}{\beta}\Big]^{d}
     \Big[ (m_\beta)^2+ (l)^2 \Big]^{\frac{d}{2}}
      \Gamma (\frac{-d}{2}) \nonumber \\
       &=& 2 \mathbb{I}_B[0;0] (2\beta) -   \mathbb{I}_B[0;0] (\beta), 
\end{eqnarray}      
where $p=2n+1$ are odd integers. But $q,l\in \mathbb{Z}^{+}$. Note that zero-mode contributions from  $\mathbb{I}_B[0;0] (2\beta)$ and $\mathbb{I}_B[0;0] (\beta)$ exactly   cancel each other.  This relation equally holds for all values of $\{k,l\},$ and we can write the general relation as 

\begin{equation}
 \mathbb{I}_F[k;l] (\beta) = 2 \mathbb{I}_B[k;l] (2\beta) -   \mathbb{I}_B[k;l] (\beta) \;\;\; \forall \; \{k;l\}.
\end{equation}

\end{document}